\documentclass[aps,pre,twocolumn,groupedaddress,showpacs,amsmath,amssymb]{revtex4}
\usepackage{graphicx}
\usepackage{subfigure}
\usepackage{epstopdf}
\usepackage{color}
\begin{document}

% Use the \preprint command to place your local institutional report
% number in the upper righthand corner of the title page in preprint mode.
% Multiple \preprint commands are allowed.
% Use the 'preprintnumbers' class option to override journal defaults
% to display numbers if necessary
%\preprint{}

%Title of paper
\title{Electrostatic interactions between discrete helices of charge}

% repeat the \author .. \affiliation  etc. as needed
% \email, \thanks, \homepage, \altaffiliation all apply to the current
% author. Explanatory text should go in the []'s, actual e-mail
% address or url should go in the {}'s for \email and \homepage.
% Please use the appropriate macro foreach each type of information

% \affiliation command applies to all authors since the last
% \affiliation command. The \affiliation command should follow the
% other information
% \affiliation can be followed by \email, \homepage, \thanks as well.
\author{Jonathan Landy}
\email[]{landy@physics.ucla.edu}
\author{Joseph Rudnick}
\email[]{jrudnick@physics.ucla.edu}
%\homepage[]{Your web page}
%\thanks{}
%\altaffiliation{}
\affiliation{Department of Physics and Astronomy, University of
California Los Angeles, Los Angeles, CA 90095-1547}

%Collaboration name if desired (requires use of superscriptaddress
%option in \documentclass). \noaffiliation is required (may also be
%used with the \author command).
%\collaboration can be followed by \email, \homepage, \thanks as well.
%\collaboration{}
%\noaffiliation

\date{\today}

\begin{abstract}
We analytically examine the pair interaction for parallel, discrete helices of charge.  Symmetry arguments allow for the energy to be decomposed into a sum of terms, each of which has an intuitive geometric interpretation.  Truncated Fourier expansions for these terms allow for accurate modeling of both the axial and azimuthal terms in the interaction energy and these expressions are shown to be insensitive to the form of the interaction.  The energy is evaluated numerically through application of an Ewald-like summation technique for the particular case of unscreened Coulomb interactions between the charges of the two helices.  The mode structures and electrostatic energies of flexible helices are also studied.  Consequences of the resulting energy expressions are considered for both F-actin and A-DNA aggregates.  
\end{abstract}

% insert suggested PACS numbers in braces on next line
% insert suggested keywords - APS authors don't need to do this
%\keywords{}

%\maketitle must follow title, authors, abstract, \pacs, and \keywords
\maketitle

% body of paper here - Use proper section commands
% References should be done using the \cite, \ref, and \label commands

\section{Introduction}

Many important biological polymers are both acidic and helical in structure.  Specific examples include DNA and F-actin, a key component of the cellular cytoskeleton.  When placed in water, each of the acidic subunits of these polymers becomes negatively charged.  This results in a discrete electrostatic charge distribution which follows the shape of the polymer's helical backbone.  Because these molecules each carry a large net negative charge, they will strongly repel one another under typical conditions.  However, through the introduction of multivalent counterions, cross-linking agents, or osmotic stress, the molecules can be condensed to a high density \cite{Gel:00, Lev:02, Won:06}.  This ability to induce aggregation of like-charged molecules is important for biological systems.  Compacted DNA is found in cell nuclei, bacteria, and virus capsids, for example, while F-actin bundles play an important role in the processes of cell motion and division \cite{Kin:01, Min:04, Sch:06,Bra:01, Bea:02}.  At high density, the interaction potential of two neighboring molecules will depend strongly on their relative positions and on the many parameters specifying their helical structure.  This sensitivity can result in various conformational phase transitions dependent upon lattice and individual helix symmetry couplings \cite{Kor:07}.  Understanding the physical mechanisms behind these transitions and how they are used for biological control has been one of the major goals of the literature considering this topic.

To model these systems many previous theoretical studies have considered the interactions between continuous helices of charge.  Such considerations led to interesting results which appear to rationalize various experimentally observed phenomena.  These include the possibility of a B to A conformational phase transition for condensed DNA bundles \cite{Kor:98:1,Rudd:07} as well as more general frustration-induced packing geometry transitions \cite{Harr:02,Gra:06}.  

Although physical systems always consist of discrete charge distributions, theoretical studies of the interaction between discrete helices of charge have been comparatively limited.  Various studies have considered particular integer number of charges per turn systems \cite{Kor:98:1, Har:03, All:00, All:04} but to the authors' knowledge, only one previous study has looked at the general discrete case \cite{Kor:98:2}.  In that reference it was demonstrated that the interaction energy of two parallel helices is a discontinuous function in the angle between charges on a single helix.  This remarkable result followed from the fact that helices with a rational number of charges per backbone turn, hereafter referred to as rational helices, may interact through modes which vanish for a pair of corresponding irrational helices.  Here, the modes considered are terms in the series representation obtained in Ref.~\cite{Kor:97:1} for the interaction between two general cylindrical charge distributions, explicitly taking into account the effects of both adsorbed and free counterions.    Modes unique to rational helices were found to allow for energy reduction.  This discrete effect could possibly play a significant role in determining the conformational twist of interacting helices.  For example, through the consideration of an idealized, perfect helical model, it was shown that this mechanism could provide a sufficient amount of energy to allow for the twisting of B-DNA from its isolated value of $10.4$ charges per turn to the $10.0$ charges per turn observed in aggregate \cite{Kor:98:2}.  Although the same authors later argued that sequence dependent variations in twist should wash out this effect for B-DNA \cite{Kor:07}, their earlier work indicates that the consequences of discreteness can be significant.  Thus, in principle, the phase diagrams of these systems should depend not only on the smoothed out helical shape of their constituent molecules, but also on the symmetries of their discrete charge distributions.

In this paper, we revisit the interaction between a pair of helical discrete charge distributions.   A simple model is considered in an effort to focus directly on the symmetries of the interaction.  These symmetries allow us to demonstrate useful relationships between the interaction energies of each of the three most commonly studied models for these systems.  These include the discrete helical, continuous helical, and cylindrical models, the latter being a model in which each molecule is replaced by a continuous cylindrical surface charge.  This approach allows for a convenient decomposition of the discrete interaction energy which takes the cylindrical interaction energy as a base energy and then adds on independent correction terms associated with different aspects of the helical shape of the molecules.  The rational discrete correction terms, equivalent to the extra modes discussed in Ref.~\cite{Kor:97:1}, are considered in some depth.  

Algebraic analysis allows us to determine information regarding the phases and amplitudes of the Fourier series expansions for each of the correction terms.  For example, an extra symmetry of the system is considered and is shown to cause a reduction in the energy benefit of rationality for certain high-symmetry orientations.  In addition, each of the series are shown to be exponentially convergent.  Truncation of these series therefore allows for simplified, approximate expressions for the energy to be obtained which capture the basic structure of the energy landscape observed numerically.   Although we focus on the electrostatic interaction, it is later shown that the same expressions also apply quite generally and can be used to model other forms of interaction.  We apply the effective energy expressions to the examples of F-actin and A-DNA to demonstrate how they may be used to make strong statements regarding the azimuthal interactions and frustrations in these systems.  As discussed previously \cite{Kor:98:1}, reduction of azimuthal frustrations may help to explain why particular lattice packing structures and twist angles are observed in experiments.  

In addition to the static helix-helix interaction energy, we also consider the mode structures of both an isolated and an interacting pair of flexible helices.  An instability is indicated by the single helix mode analysis for certain choices of the angle between charges.  To understand this, we briefly consider the electrostatic energy of a single helix of charge.  For a pair of interacting helices, we show that a gap is expected at long wavelengths between the oscillation frequencies of a rational system and those of a corresponding irrational system.  Appropriate neutron scattering experiments may thus provide information regarding the rationality of twist angles in bundled sets of helices.  In principle, such experiments could directly test whether sequence dependent twist variations indeed wash out the effects of rationality in B-DNA systems.  This was argued to be the case by the authors of Ref.~\cite{Kor:07} on the basis of angle variations inferred from x-ray diffraction data.  Angle variation need not only result from base-pair sequence variation, however.  Local twisting should also occur within these systems, resulting in an increase in the energy benefit associated with discreteness.  This suggests that further study of disordered, flexible systems is required in order to rule out discreteness as the cause of twisting in B-DNA aggregates and elsewhere.  The present study provides a first step in this direction.

The paper proceeds as follows: the mathematical formulation for the problem considered is presented in the following section; in section III geometric symmetry arguments are presented which provide the basis for the energy decomposition; section IV addresses the effects of symmetry on the phases and amplitudes of the Fourier components of the terms in the decomposition; section V contains the energy and mode analysis for azimuthally flexible systems;  section VI explores the possible consequences of our simplified energy expressions for physical F-actin and A-DNA aggregates;  section VII contains concluding remarks; finally, an outline of the Ewald summation technique applied to obtain quick numerical evaluations of the energy appears in an appendix.

\begin{figure}\scalebox{0.2}
{\includegraphics{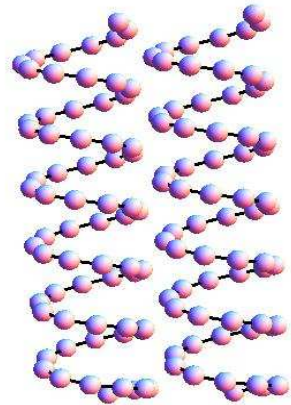}}
\caption{\label{fig:start} The fundamental configuration considered.}
\end{figure}

\section{Mathematical formulation and numerical evaluation}
\begin{figure}\scalebox{0.6}
{\includegraphics{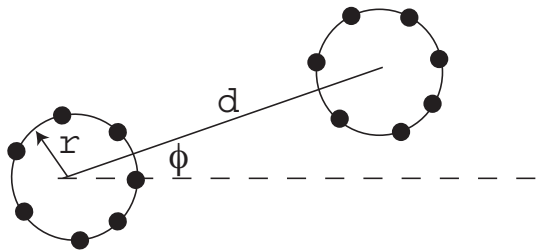}}
\caption{\label{fig:end} The helical charge distributions, shown end-on.}
\end{figure}

The fundamental configuration of charges considered is depicted in Fig.~\ref{fig:start}.  Two identical, infinite helices of discrete charge lie parallel to one another, each with radius $r$ and separated by a distance $d$.  The interaction energy of the two helices may be formally expressed as
\begin{eqnarray}
\label{E:sum:def}
E =  \sum_{\theta_1, \theta_2} \frac{\exp[-a_s R]}{R},
\end{eqnarray}
where the sum is over all pairs of charge, one taken from each of the two helices.  Here we have assumed a Yukawa type individual charge potential, consistent with the Debye-H\"uckel screening approximation \cite{Deb:23}, $a_s$ is the screening parameter, and the distance $R$ between two charges is given by
\begin{eqnarray}\nonumber
R^2 &=& [d- r\cos(\theta_1 -\phi) + r \cos(\theta_2 - \phi)]^2  \\ \nonumber
&& + [r \sin(\theta_1 - \phi) - r \sin(\theta_2- \phi)]^2  + a^2 [\theta_1 - \theta_2 + \zeta]^2 \\ \nonumber
&=&
d^2 - 2rd[\cos(\theta_1 - \phi) - \cos(\theta_2 - \phi)] \\ 
&&+ 2r^2[1- \cos(\theta_1 - \theta_2)] + a^2[\theta_1 - \theta_2 + \zeta]^2.
\label{R:def}
\end{eqnarray}
In the above, the parameters $\theta_1$ and $\theta_2$ specify the azimuthal angles of the two charges being summed over, $\phi$ is defined in Fig.~\ref{fig:end}, $a$ is related to the helical pitch, and $\zeta$ describes a vertical or axial shift between the two helices.  To begin, we assume perfect helices and the angles $\theta_1$ and $\theta_2$ are written as
\begin{eqnarray}
\label{n1:def}
\theta_1 &=& n_1 \psi \\
\label{n2:def}
\theta_2 &=& n_2 \psi + \Delta \psi,
\end{eqnarray}
where the $n_i$ are integers to be summed over, $\psi$ is the azimuthal angle between adjacent charges on a single helix, and $\Delta \psi$ is the azimuthal shift of the charges on the second helix relative to those on the first.  For rational $\psi$ we write
\begin{eqnarray}
\label{psi:def}
\psi = 2 \pi \frac{m_{\psi}}{n_{\psi}},
\end{eqnarray} 
with $m_{\psi}$ and $n_{\psi}$ relatively prime.  Equation (\ref{psi:def}) indicates that each helix has $n_{\psi}$ charges for each $m_{\psi}$ turns of the backbone.  It follows that the discrete energy is, in general, periodic in $\zeta$ with period $2 \pi m_{\psi}$, in $\Delta \psi$ with period $\psi$, and in $\phi$ with period $2 \pi/n_{\psi}$.

All numerical values of the energy presented here are for the case of Coulomb interactions with no screening.  That is, $a_s$ was set to zero.  This was done to simplify the expressions being summed over; while this somewhat removes the model from the physical systems considered, the numerically calculated energies retain the symmetries of these systems.  It is these symmetries which are the focus of study in this paper, and fortunately, these are quite insensitive to the individual potential functions used.  The following two sections demonstrate how these symmetries allow for a characterization of the interaction energy.  Generalization of these results will be discussed in section VI.

\section{Geometric symmetry arguments}

\begin{figure*}
\subfigure[\label{fig:rot1}]{
\includegraphics[scale=0.6]{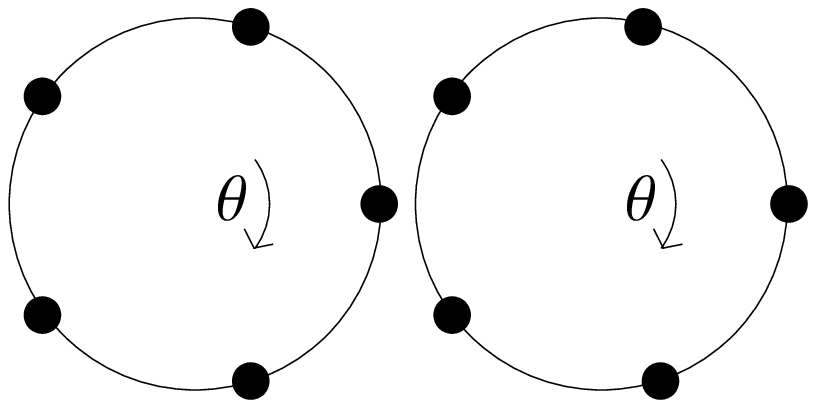}
}
\subfigure[\label{fig:rot2}]{
\includegraphics[scale=0.6]{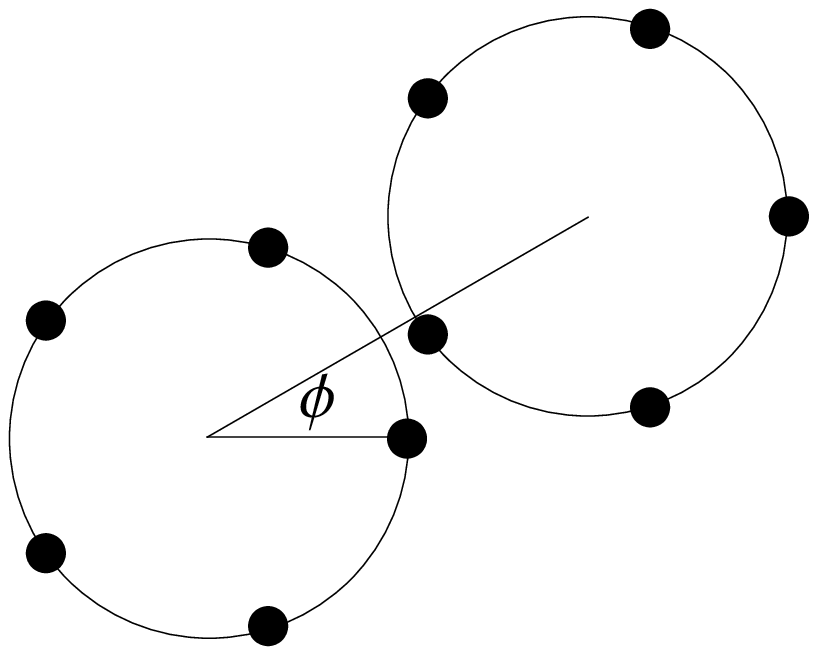}
}

\caption{ \ref{fig:rot1} Because of angular drift, the energy for an irrational pair will be given by the $\theta$ average of a nearby rational pair.   \ref{fig:rot2}  Taking the first helix's perspective of the $\theta$ average, we see this is equivalent to a $\phi$ average, with $\Delta\psi$ and $\zeta$ kept fixed.  In the figure $\phi
=\theta$.}
\end{figure*}

Consider the potential felt by a charge on one irrational helix due to the charges on a second, identical helix.  Locally the second irrational helix will look like a helix with a nearby rational number of charges per turn.  Since the potential of the second helix is a continuous function in $\psi$, it follows that the potential energy of the first charge is the same as it would be if the second helix were rational.  However, after many turns the angles at which the discrete charges on the two helices are placed will begin to drift with respect to
the corresponding positions for a rational pair of helices.  This is because the $\psi$ value for the irrational case is very close but not equal to the rational $\psi$ value.  It follows that the irrational interaction energy is the $\theta$ average of the rational interaction energy.  See Fig.~\ref{fig:rot1}.

We can look at this averaging from another perspective, keeping $\theta_1$ fixed and allowing the other variables to adjust.  It is easy to see that through this averaging $\phi$ will rotate through $2 \pi$ radians while $\Delta\psi$ and $\zeta$ will remain fixed throughout. See Fig.~\ref{fig:rot2}.  It follows that the energy for an irrational pair is the $\phi$ average of a nearby rational pair with the same parameter values for $\Delta \psi$ and $\zeta$.

If we take the $\Delta \psi$ average of the irrational energy we obtain the interaction energy between one continuous helix and a second irrational helix.  This is equivalent to the interaction energy between two continuous helices, however, because the set $\{ k \psi: k \in \mathbb Z \}$ is dense modulo $2\pi$ for irrational $\psi$.  We may, therefore, smear out the charge over the first helix's backbone as well without changing the energy.  It follows that the continuous pair interaction energy is the $\Delta \psi$ average of the irrational pair energy.

Finally, taking the $\zeta$ average of the continuous case, we obtain the interaction energy between a continuous helix and a cylinder with charge uniformly distributed over its surface.  The potential due to the cylinder is independent of the position along the direction of its axis.  The energy is thus unaffected when the continuous helix is averaged out in this direction, as well, and so this is equivalent to the interaction between two cylinders of charge.  The cylindrical model interaction energy may, therefore, be obtained from a $\zeta$ average of the continuous system's energy.

One significant consequence of the above follows from the fact that if one system's energy is the average of another's, the latter's energy can take on values both larger and smaller than that of the former's through an appropriate choice of the parameter being averaged over.  Physically, both the irrational and rational situations may be realized.  As the irrational case is the $\phi$ average of the rational energy, it follows that an interacting pair of helices can often decrease its interaction energy through an adjustment of the angle $\psi$ to a nearby rational value.  In the following section we demonstrate that the amplitude of the $\phi$ dependence decreases exponentially with the value of $n_{\psi}$.  This statement specifies the manner in which the rational energy benefit depends on the rationality of $\psi$.  With this information we can now determine whether or not all irrational helices will be induced to twist when interacting with a second identical helix.  

\begin{figure*}
\subfigure[\label{fig:stair1}]{
\includegraphics[scale=0.75]{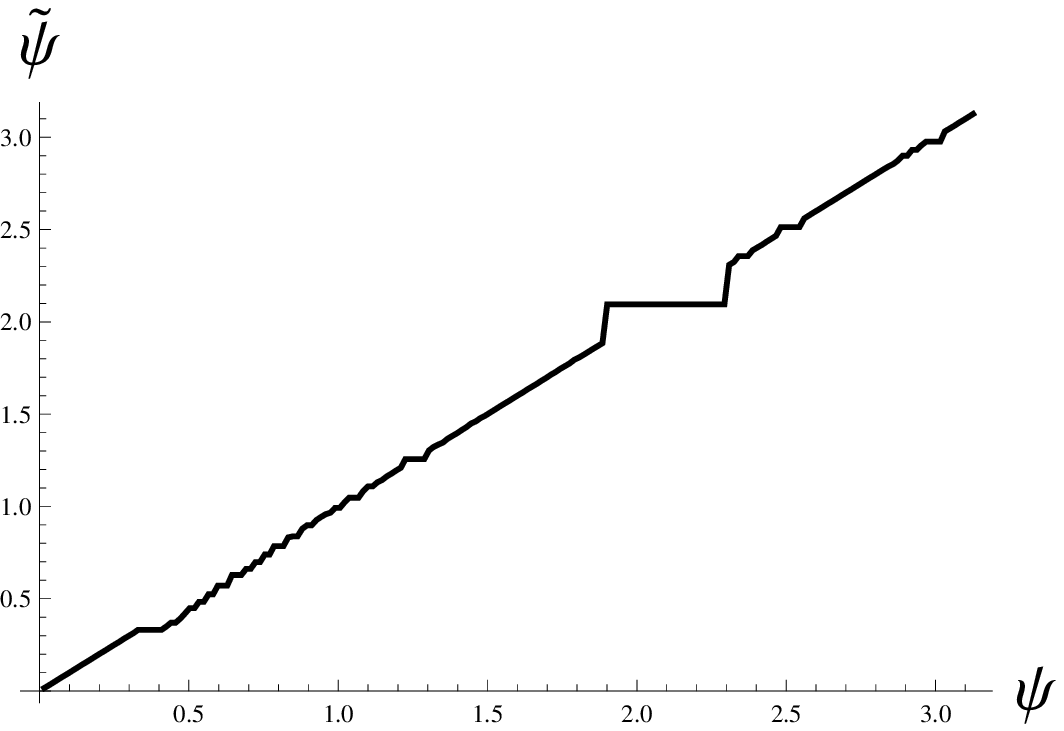}
}
\subfigure[\label{fig:stair2}]{
\includegraphics[scale=0.75]{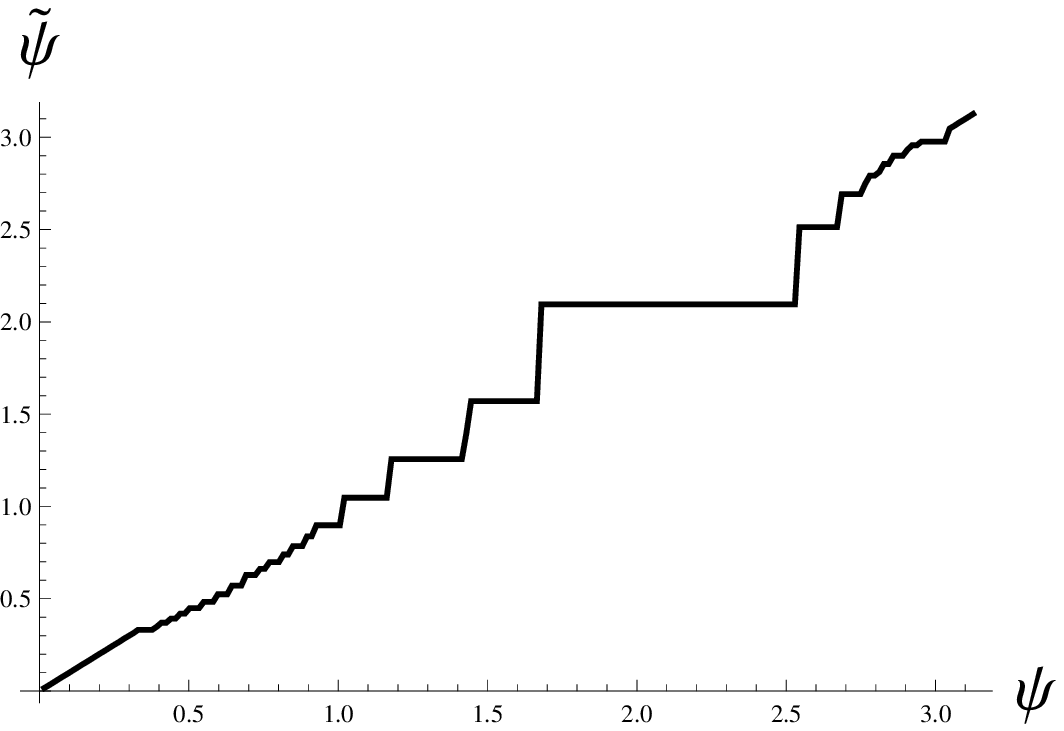}
}

\caption{Shown are two plots of the preferred angle between charges $\tilde{\psi}$ for an interacting pair of helices versus the value of $\psi$ assumed for an isolated helix.  The helices were given an arbitrary torsional spring constant which resists twisting, which is assumed to occur in such a way that the linear charge densities of the helices is fixed.  In \ref{fig:stair1}, the separation distance is large enough so that only an incomplete devil's staircase is observed.  In  \ref{fig:stair2}, the separation has been decreased and a filled in staircase is observed.  Note that the one charge per turn system was excluded.  If this is retained, it dominates the small $\psi$ region of the plot and all irrational helices are observed to twist.
}
\end{figure*}

We start by positing that the energy gain associated with a given rational helix pair scales as $\exp[- \gamma n_{\psi}]$.  Further, we suppose that this energy gain is sufficient to twist all irrational helix pairs in the twisting range $(\psi - \delta \psi, \psi + \delta \psi)$ to the rational value $\psi$.  It follows that $\delta \psi$ is also proportional to $\exp[- \gamma n_{\psi}]$.  If we sum up the widths of the twisting regions around each rational $\psi$ value less than $2\pi$, we get

\begin{eqnarray} \nonumber
W(\gamma) &\equiv & \sum_{m/n \in  \{ \mathbb{Q}<1 \} } \exp[- \gamma n] \\ \nonumber
&< & \sum_{n=1}^{\infty}\sum_{m=1}^{n} \exp[- \gamma n] \\
&= & \frac{e^{\gamma}}{(e^{\gamma}-1)^2}.
\end{eqnarray}
In the second line we have replaced the sum over all rational values by the sum over all relevant integer denominators and numerators.  Note that this significantly over-counts the number of rationals since we are including terms which are not in reduced form. Additionally, we overestimate the portion of the interval associated with rational helices, in that some of the basins of attraction of rational helices are ``shadowed" by others.   In spite of this over-counting, the sum over the twisting widths is bounded.  Indeed, for large separations, we expect $\gamma$ to be large.  In this case $W(\gamma)<2 \pi$, implying that not all the irrational $\psi$ values in $(0, 2\pi)$ will be twisted.  However, as the helices approach one another, more and more irrational helices should experience a $\phi$ induced twist since the energy benefit of a rational $\psi$ value increases with proximity.  This expectation is confirmed when one plots the optimal angle $\tilde{\psi}$ for an interacting pair of helices versus the isolated $\psi$ value at different spacings $d$.  This is done in Figs.~\ref{fig:stair1} and \ref{fig:stair2}.  At large $d$ one observes an incomplete devil's staircase.  As the separation distance is decreased, a filled in staircase is observed and nearly all helices are twisted to a nearby rational value.

A second significant consequence of the above averaging arguments is that they often allow for a convenient decomposition of the interaction energy, as mentioned above.  Beginning with the cylindrical model interaction energy, we may add on corrections for the continuous helix, discrete irrational, and finally discrete rational terms.  Each new correction term adds dependence to the energy on a new parameter.  In addition, at the level at which energy dependence on a given parameter is first introduced, the energy has a well defined periodicity in that parameter.  For example, although the irrational energy is not periodic in $\zeta$, the continuous energy is and has the finite period of $2\pi$.  The periodicity at each level allows each of the correction terms to be expanded in a Fourier series.  Further, except in certain extreme limits, these Fourier expansions are quickly damped.  Truncation of these series thus allows for simple, approximate expressions for the energy to be obtained which are consistent with the sinusoidal forms typically observed numerically.  While first order truncations are often sufficient, higher order harmonics may be required to accurately model the energy for parameter locations which allow for close charge interactions.  This is often the case for $\zeta$ values near $\pi$ and for small $d$ values, for example.   In this case the helical backbones are close to one another and small adjustments in $\Delta \psi$ or $\phi$ may allow for large increases in the energy.  In practice, physical systems should often be found far from such parameter locations, however, and first order approximations for the energy are therefore acceptable.  

Numerical observations have indicated that the continuous energy often dominates both the $\Delta \psi$ and $\phi$ dependence by at least one order of magnitude.  In this case, we say that the system is in the continuous limit and think of the rational and irrational terms as perturbative correction terms for the energy.  For small values of the parameter $d$, however, the correction terms can also have amplitudes on the order of magnitude of the thermal $k_BT$ energy scale per azimuthal persistence length \cite{Kor:98:1}.  Notable extreme limits where the continuous limit is not valid include the large $a$ limit, in which the charges on each helix are separated by large axial distances, and the small $a$, rational limit, where each helix looks much like a grouping of $n_{\psi}$ lines of continuous charge.  Most physical systems appear to be somewhere between these two limits, however, and the energy may be considered to be in the continuous limit with first or second order sinusoidal discrete correction terms sufficient.  

\section{Phases and amplitudes}
To continue the characterization of the pair interaction energy, we now consider the phases and amplitudes of the Fourier expansions of the correction terms described above.  In addition, we briefly discuss the irrational energy's $\psi$ dependence.  This $\psi$ dependence is of interest since it may provide a mechanism for discrete interaction induced twisting for some systems.  

\subsubsection*{Continuous $\zeta$ dependence}
The phase of the continuous $\zeta$ dependence may be determined by explicitly writing down the energy in integral form and differentiating to find extrema.  Doing this we find
\begin{eqnarray}
\frac{d E_c}{d \zeta} = \int \int \frac{d}{dx}(\frac{\exp[-a_s x]}{x}) \frac{d x}{d \zeta} d \theta_1 d \theta_2 .
\end{eqnarray}
Here,
\begin{eqnarray} \nonumber
\frac{d x}{d \zeta} &=& \frac{d x}{d x^2}\frac{d x^2}{d \zeta} \\
&\propto & a^2 (\theta_1 - \theta_2 + \zeta),
\end{eqnarray}
and
\begin{eqnarray}\nonumber
\label{xsqrcont}
x^2 = d^2 + 2 r^2 + a^2 (\theta_2 - \theta_1 + \zeta)^2 - 2 r d( \cos \theta_1 - \cos \theta_2)\\ - 2 r^2 \cos (\theta_1 - \theta_2).
\end{eqnarray}
The integrand is odd about $\theta_1 =\zeta$ and $\theta_2= 0$ when $\zeta = k \pi$, with $k$ an integer.  The continuous energy correction term may therefore be expanded as
\begin{eqnarray}
\label{energy:zeta}
E_{\zeta} = \sum_{k=1}^{\infty}A_{k,\zeta}\cos(k \zeta).
\end{eqnarray}
The magnitudes of the coefficients $A_{k, \zeta}$ have been considered in previous studies of the continuous energy and have been proven to decay exponentially with both $d$ and $k$ \cite{Kor:97:1}.  The decay rate was found to be $\exp[-2 \pi \sqrt{k^2 + a_s^2}d/P]$, where $P$ is the pitch of the helices.   We note that this result could also be obtained through a superposition of  interactions between lines of periodically spaced point charges.  The rapid convergence of the above sum leads us to expect the first coefficient $A_{1,\zeta}$ to take on a negative value quite generally.

\subsubsection*{Irrational $\Delta \psi$ dependence}

\begin{figure}
\includegraphics[scale=0.8]{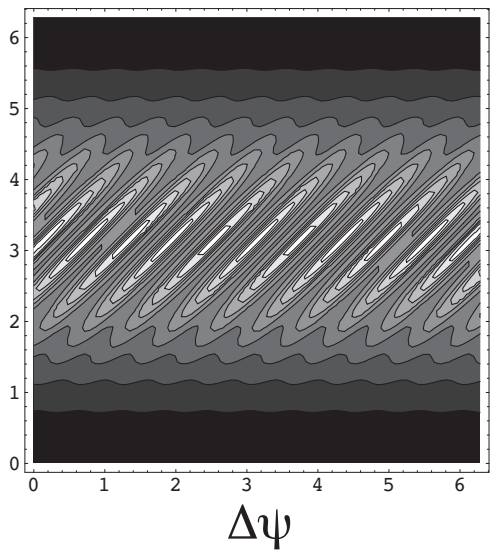}
\caption{\label{fig:irrat} A contour plot of the irrational energy versus $\Delta \psi$ and $\zeta$.}
\end{figure}

Explicit integral expressions for the Fourier coefficients of the irrational energy may be obtained as follows. Recalling that the set $\{ k \psi: k \in \mathbb Z \}$ is dense modulo $2\pi$ for irrational $\psi$, we set $n_1 = n_2 + m_a$ in Eq.~(\ref{n1:def}).  The sum on $n_2$ may then be replaced by an integral over $0$ to $2 \pi$.   Next we apply the Poisson sum rule to the sum on $m_a$ and obtain for the energy per charge on the second helix
\begin{eqnarray}
\label{engy:sim:cont:new}\nonumber E_{I} = \frac{q^2}{2 \pi
\psi}\int_{-\infty}^{\infty} \int_0^{2\pi}\frac{e^{-a_s
x}}{x}d\theta_1 d\theta_2 \sum_k e^{2\pi i k (\frac{\theta_2 -
\theta_1 - \Delta \psi}{\psi})}, \\
\end{eqnarray}
where $x^2$ is again given by Eq.~(\ref{xsqrcont}).  Note that the $k=0$ term above gives the continuous energy.  Upon averaging over $\Delta \psi$ the other terms vanish, an observation consistent with the above geometric averaging result.

Parity arguments again allow us to determine the extrema of the irrational $\Delta \psi$ dependence, but only when $\zeta$ is an integer multiple of $\pi$.   At integer $\zeta / \pi$, it may be easily shown from Eq.~(\ref{engy:sim:cont:new}) that there are irrational $\Delta \psi$ extrema at $\Delta \psi = \zeta, \zeta \pm \psi/2, \zeta \pm 2 \psi/2, ...$, etc.  Although symmetry arguments alone are insufficient to determine the phases away from integer $\zeta /\pi$, numerical observations indicate that this phase is often quite linear in $\zeta$.  See Fig.~\ref{fig:irrat}.  The slope of this phase dependence in $\zeta$ depends upon the parameters $a$, $r$, and $d$, however.  This information allows us to expand the irrational energy as 
\begin{eqnarray}
\label{energy:deltapsi}
\nonumber
E_{\Delta \psi} = \sum_{k=1}^{\infty}A_{k,\Delta \psi}\cos( \frac{2\pi k}{\psi}[\Delta \psi -(1- \frac{s \psi}{2 \pi}) \zeta + O(\zeta ^2)]),\\
\end{eqnarray}
where $s$ is some integer and the $O(\zeta^2)$ terms in the phase must vanish whenever $\zeta/\pi$ is integer.

We may demonstrate that the $A_{k,\Delta \psi}$ in Eq.~(\ref{energy:deltapsi}) decay exponentially with $k$ by returning to the integral representation in Eq.~(\ref{engy:sim:cont:new}).  Changing variables to $(u,v) \equiv (\theta_2 + \theta_1, \theta_2 - \theta_1)$, it may be shown that for $d>2r$, one may always add to $v$ a positive, finite imaginary part $i \kappa$ without crossing any singularities.  For large $d$, $\kappa$ is bounded by $d/a$ and
\begin{eqnarray}
A_{k,\Delta \psi} \propto \exp[-\frac{2 \pi k d}{a \psi}],
\end{eqnarray}
also consistent with the discrete lines of charge limit.

\subsubsection*{Rational $\phi$ dependence}
Once again setting $n_1 = n_2 + m_a$ in Eq.~(\ref{n2:def}), the rational energy per charge on the second helix may be expressed as 
\begin{eqnarray}
\label{rat:sum} E =\frac{q^2}{m_{\psi}}
\sum_{m_a=-\infty}^{\infty}\sum_{k=1}^{n_{\psi}} \frac{e^{-a_s x}}{x},
\end{eqnarray}
where
\begin{eqnarray}
\label{xsq:gen} \lefteqn{x^2 = d^2 + 2r^2 + a^2(m_a \psi -\Delta\psi + 
\zeta)^2 } \nonumber \\&& + 4rd\sin(\frac{m_a \psi + \Delta \psi}{2} -\phi +\frac{2\pi k}{n_{\psi}})  \sin(\frac{m_a \psi - \Delta \psi}{2}) \nonumber \\ 
&& -2r^2\cos(m_a\psi-\Delta\psi).
\end{eqnarray}
It is easy to see that the above is even in $\phi$ about the point $(\Delta \psi - \pi)/2$, independent of $\zeta$.  To show this, one need only note that $m_a \psi /2 = m_a m_{\psi} \pi/n_{\psi}.$  This phase shift will not affect the parity of inside sum over $k$ since it always shifts the sum by an integer multiple of $\pi/n_{\psi}$.  It follows that the rational energy correction term may be expressed as
\begin{eqnarray}
\label{energy:rational}
E_{\phi} = \sum_{k=1}^{\infty} A_{k,\phi} \cos[n_{\psi}k (\phi - \frac{\Delta \psi - \pi}{2})].
\end{eqnarray}

To examine the convergence of the sum in Eq.~(\ref{energy:rational}), we again focus on the inside sum of Eq.~(\ref{rat:sum}).  Letting $\delta = \phi-\frac{m_a \psi + \Delta \psi}{2} $, we rewrite this inside sum as
\begin{eqnarray}
\label{f:one}
f(n_{\psi},\phi) \equiv \sum_{k = 1}^{n_{\psi}} g(a + b \cos(\frac{2 \pi k}{n_{\psi}} - \delta)).
\end{eqnarray}
Here $a$ and $b$ are constants over the $k$ sum, with $b<a$ since $d> 2r$, and $g$ is the individual charge potential.  To obtain an expression for the first coefficient in Eq.~(\ref{energy:rational}), we Taylor expand the function $g$ about the point $a$
\begin{eqnarray}\label{g:one}
g(a+x) = \sum_j a_j x^j,
\end{eqnarray}
and note that only those $j\geq l n_{\psi}$ contribute to $A_{l,\phi}$.  This is because to get an argument containing $l n_{\psi} \delta$, you need a product containing at least $l n_{\psi}$ factors of $\cos (\delta)$.  
Plugging in the Coulomb potential and summing on $j\geq l n_{\psi}$ gives
\begin{eqnarray} \label{f:four}\nonumber
\sum_{j = 0}^{\infty} \frac{g^{(l n_{\psi}+j)}(a)}{(l n_{\psi}+j)!}b^{l n_{\psi}+j}\sum_{k = 1}^{n_{\psi}} \cos(\frac{2 \pi k}{n_{\psi}} - \delta)^{l n_{\psi}+j}  \propto (\frac{b}{a})^{l n_\psi}.\\
\end{eqnarray}
It may be shown that the fluctuating portion in the cosine sum in Eq.~(\ref{f:four}) is of order unity.  Since $b<a$, it follows that the series Eq.~(\ref{energy:rational}) is exponentially convergent and that the decay rate is proportional to $n_{\psi}$.  A careful look at the values of $b$ and $a$ above shows that the $\phi$ dependence decays algebraically with $d$.  Thus, at large distances, the rational correction terms will dominate the continuous helix correction terms.  Typically, this will occur at sufficiently large distances that each of these corrections are effectively negligible.  For small $a$ this limit may become significant, however.

\begin{figure*}
\subfigure[\label{fig:ratA}]{
\includegraphics[scale=0.5]{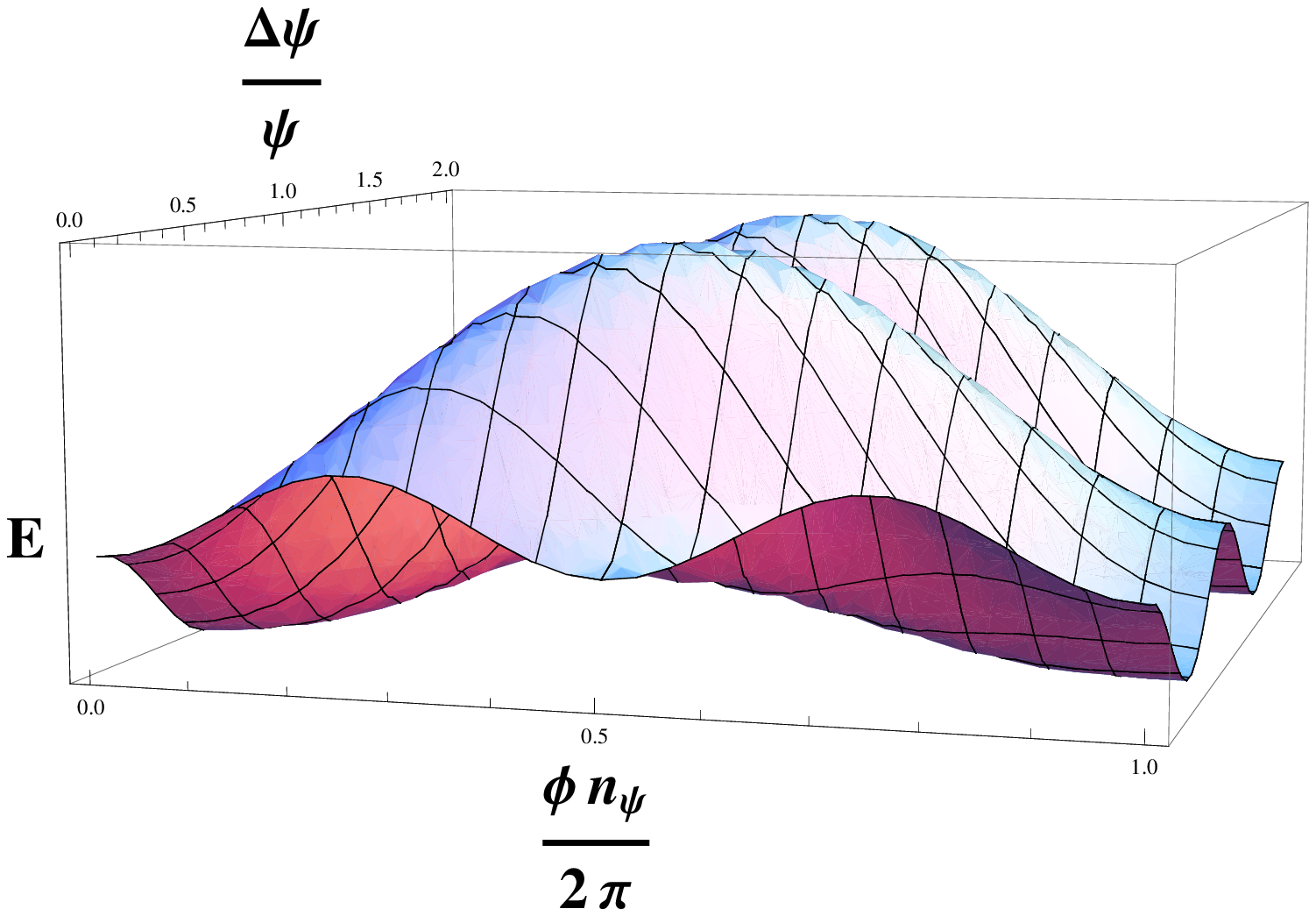}
}
\subfigure[\label{fig:ratB}]{
\includegraphics[scale=0.55]{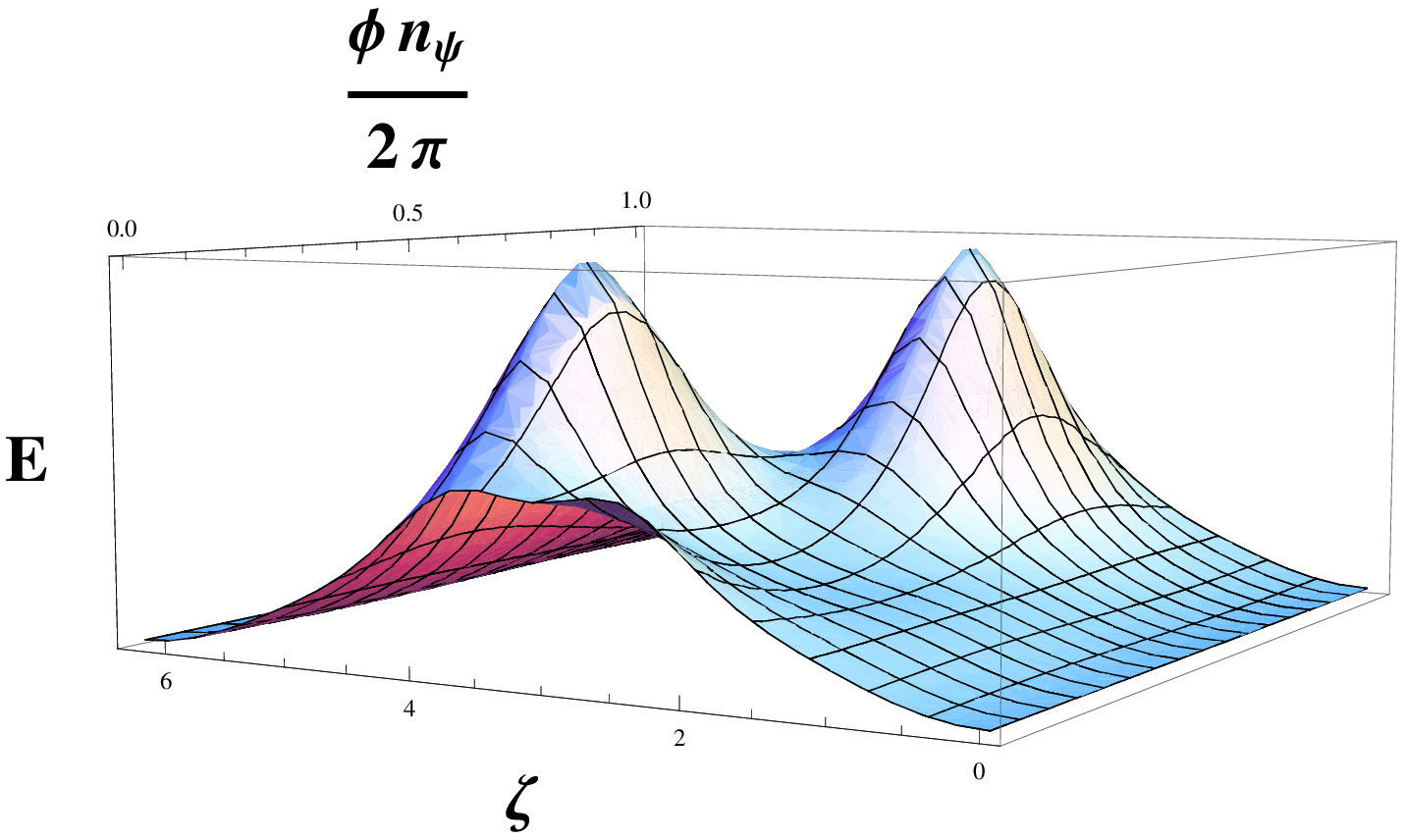}
}
\caption{\ref{fig:ratA} A plot of the rational energy versus $\phi$ and $\Delta \psi$ for $\psi = 2 \pi/7$ and $\zeta = 0$.  Note that the first $\phi$ component vanishes when $\Delta \psi$ is an integer multiple of $\psi$, consistent with Table \ref{table1}.  \ref{fig:ratB} A plot of the rational energy versus $\phi$ and $\zeta$ for $\psi = 2 \pi/5$ and $\Delta \psi =0$.  While the phase of the $\phi$ dependence is independent of $\zeta$, the amplitude is not and is observed to change signs as $\zeta$ moves through $2\pi$.}
\end{figure*}

It turns out that the phase dependence in Eq.~(\ref{energy:rational}) may sometimes have an interesting impact on the amplitude of the coefficients.  Consider what happens when $\phi$ and $\Delta \psi$ are each adjusted by one period.  Plugging into the $k=1$ term in Eq.~(\ref{energy:rational}) gives
\begin{eqnarray} \nonumber
 A_{1,\phi}(\Delta \psi +  \psi, \zeta) \cos[n_{\psi} (\phi + \frac{2 \pi}{n_{\psi}} - \frac{\Delta \psi + \frac{2 \pi m_{\psi}}{n_{\psi}} - \pi}{2})] \\ \nonumber
= (-1)^{m_{\psi}} A_{1,\phi}(\Delta \psi +  \psi, \zeta) \cos[n_{\psi} (\phi  - \frac{\Delta \psi  - \pi}{2})]. \\
\end{eqnarray}
From periodicity, the energy must be unaffected by this shift in $\phi$ and $\Delta \psi$.  It follows that

\begin{eqnarray}
A_{1,\phi}(\Delta \psi +  \psi) =  (-1)^{m_{\psi}} A_{1,\phi}(\Delta \psi).
\end{eqnarray}
Therefore, for $m_{\psi}$ odd, $A_{1,\phi}$ must change sign as $\Delta \psi$ is adjusted through $\psi$ radians.  Since $A_{1,\phi}$ is a smooth function in $\Delta \psi$, it follows that for odd $m_{\psi}$ there is a $\Delta \psi$ value at which the amplitude of first $\phi$ component vanishes.  This is significant because the second $\phi$ component is in general exponentially smaller than the first component.  The result is that there is little $\phi$ dependence at this value of $\Delta \psi$.  See Figs.~\ref{fig:ratA} and \ref{fig:ratB}.

In general, the particular $\Delta \psi$ location where the first $\phi$ component vanishes depends on the values taken by the other parameters.  For certain high symmetry $\zeta$ values, however, the exact $\Delta \psi$ values may be determined and are independent of the remaining parameters.   To determine these locations, we must note that the energy is invariant under the operation
\begin{eqnarray}\label{pair:symmetry}
(\phi, \Delta \psi, \zeta) \rightarrow (\phi + \pi - \Delta \psi, - \Delta \psi, - \zeta).
\end{eqnarray}
This operation is equivalent to relabeling the helices one and two.  If the first $\phi$ component vanishes, this suggests that we look for parameter locations where the period of the $\phi$ dependence is halved. That is, we search for parameter locations where the energy is unaffected when $\phi$ is adjusted by $\pi/n_{\psi}$.  Plugging into Eq.~(\ref{pair:symmetry}) we look for solutions to
\begin{eqnarray}
\label{pair:symmetry2}
E(\phi_0 , \Delta \psi,\zeta) &=& E(\phi_0 + \pi/n_{\psi},  \Delta \psi, \zeta)  \\ \nonumber 
&=& E(\phi_0 + \pi/n_{\psi} + \pi -\Delta \psi, - \Delta \psi, - \zeta).
\end{eqnarray}
The solutions to Eq.~(\ref{pair:symmetry2}) are shown in Table \ref{table1}.  Note that even $m_{\psi}$ solutions exist as well.  While the amplitude does go to zero at these even $m_{\psi}$ locations, it does not change signs.  Further, while the amplitude of the first $\phi$ component is required to vanish at other $\zeta$ values for odd $m_{\psi}$, the even $m_{\psi}$ solutions noted in Table \ref{table1} are the only solutions observed numerically.  This is consistent with our expectations: while the zero-frequency component of the Fourier expansion of $A_{1,\phi}$ must vanish for odd $m_{\psi}$, symmetry conditions do not require this to be the case for even $m_{\psi}$.  The expansion of $A_{1,\phi}$ for odd $m_{\psi}$ is discussed later in our consideration of A-DNA.

\begin{table}
\caption{
\label{table1} Solutions to Eq.~(\ref{pair:symmetry2}) which specify exact parameter locations where first $\phi$ component's amplitude is zero.  Here, the $k_i$ are arbitrary integers.}
\begin{tabular}{llll}
\hline
$m_{\psi}$ & $n_{\psi}$ & $\zeta$  & $\Delta \psi$\\ \hline

	odd & odd & $0 \pm  k_1 \pi  m_{\psi}$ & $0 \pm k_2 \psi$ \\  
	odd & even & $0 \pm  k_1 \pi  m_{\psi}$& $\frac{\psi}{2} \pm k_2 \psi$  \\ 
	even & odd & $0 \pm  k_1 \pi m_{\psi}$ & $0 \pm k_2 \psi$ or  $\frac{\psi}{2} \pm k_3 \psi$  \\  
\hline
\end{tabular}
\end{table}

\subsubsection*{Irrational $\psi$ dependence}

Due to the properties of the rational $\phi$ dependence, it is clear that the pair interaction energy is a nowhere continuous function in $\psi$.  In addition to rational energy terms there is also a direct irrational $\psi$ dependence.  This irrational $\psi$ dependence provides another possible mechanism which may be responsible for the observed twistings of helical macromolecules in aggregate.  

The functional form of the $\psi$ dependence will depend on how the helices twist or untwist.  In the case of F-actin, the heights between charges remain roughly fixed during twisting \cite{Ege:82}.  This implies that the linear charge density of the molecules is unaffected by the twisting.  Numerical plots demonstrate that for systems which twist in this way, the locations of the irrational energy minima in $\psi$ depend strongly on the parameters $d/r$ and $a/r$.  Therefore, to determine whether this mechanism is a possible cause for twisting, one must numerically examine the appropriate phase space region for the system of interest.

\subsubsection*{Resulting energy expression}
The general energy expression, then, is given by adding to the energy of two interacting cylinders of charge the expressions in Eqs.~(\ref{energy:zeta}), (\ref{energy:deltapsi}), and (\ref{energy:rational}).  It is important to remember that the $A_{k,\phi}$ depend on all of the parameters except $\phi$, the $A_{k,\Delta \psi}$ depend on all the parameters except for $\phi$ and $\Delta \psi$, and that the $A_{k,\zeta}$ depend on all parameters except for $\phi$, $\Delta \psi$, and $\zeta$.  The rule of thumb is that a particular amplitude will be large when close interactions may be introduced through the particular parameter's adjustment.  Although we focused on locations where the first component of the $\phi$ dependence vanishes, the $\Delta \psi$ dependence amplitudes can also take on both positive and negative values depending on the values of the parameters $a$ and $d$.  In general, we must resort to numerics to determine which terms dominate the energy landscape for a given system and what the signs are for the amplitudes of the various terms.

\section{Mode analysis}
We turn now to a consideration of helices which are not rigid but instead have some internal degrees of freedom.  We first consider the modes of a system of mobile charged particles constrained to move on the surface of an isolated cylinder.  Each particle has the same charge and the axis of the cylinder contains a compensating line of charge of the opposite sign which ensures a net charge neutrality for the system.  We assume an initial helix distribution for the charges, which by symmetry, is clearly stable with respect to the motions of any single charge in the system.  

\begin{figure*}
\subfigure[]{
\includegraphics[scale=0.75]{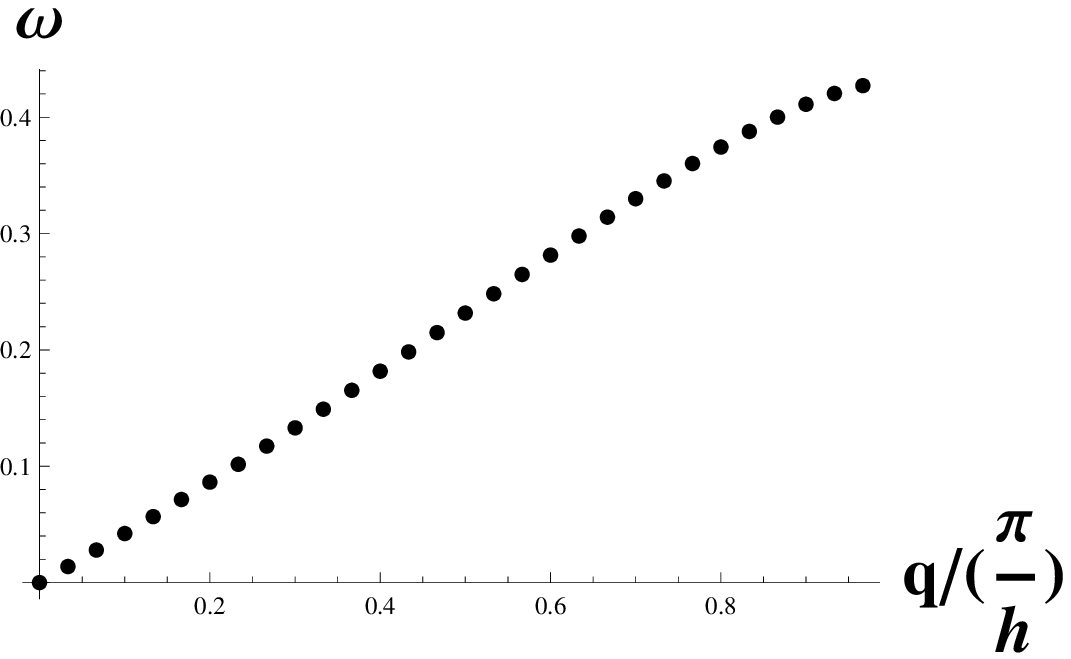}
}
\subfigure[]{
\includegraphics[scale=0.75]{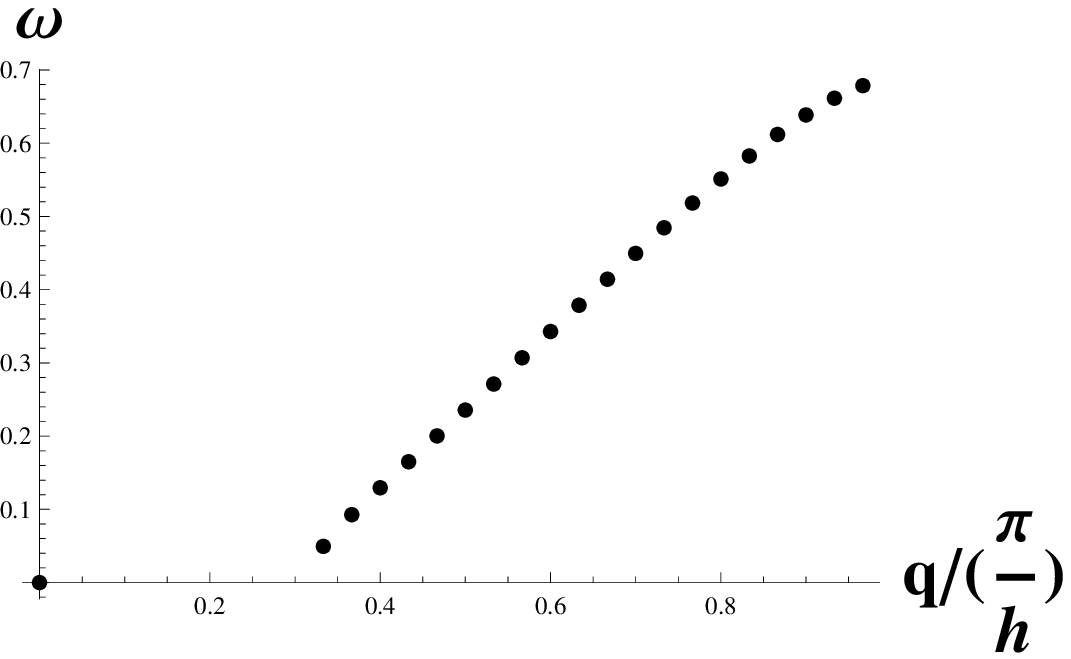}
}
\caption{\label{1mode:fig} Dispersion relation plots for the helix modes of two $\psi=\pi$ systems.  In the first $h=3.0 r$ and in the second $h=1.8 r$.  In this second case, the frequencies for certain long wavelength modes were found to be complex indicating an instability. }
\end{figure*}

For simplicity, we consider modes in which the charges are only allowed to rotate in the azimuthal direction.  Thus, the axial positions of the charges are fixed.  Recall that this is roughly how the charges fluctuate in F-actin systems.  For a given mode, the position of the charges can be written as
\begin{eqnarray}
\vec{r}_n = \vec{R}_n + \vec{u}_n,
\end{eqnarray}
where $\vec{R}_n$ describes the equilibrium position of the charge $n$.  This is given by
\begin{eqnarray}
\vec{R}_n = r (\hat{x} \cos n\psi + \hat{y} \sin n\psi) + \hat{z}h n,
\end{eqnarray}
where we have introduced the notation $h=a \psi$ for the axial rise per charge.  The vector $\vec{u}_n$ is the displacement from equilibrium and is given to second order in the mode amplitude $A_q$ by
\begin{eqnarray}\nonumber
\lefteqn{\vec{u}_n= (A_q e^{i q n h}+A_q^* e^{-i q n h})(-\hat{x} \sin n\psi + \hat{y} \cos n\psi)}\\ &&- \frac{(A_q e^{i q n h}+A_q^* e^{-i q n h})^2}{2 r}(\hat{x} \cos n\psi + \hat{y} \sin n \psi).
\end{eqnarray}
We assume a Coulomb interaction between charges and make use of the identity
\begin{eqnarray}\label{CoulInt}
r^{-1} = \pi^{-1/2}\int_0^{\infty} w^{-1/2}e^{-w r^2} dw.
\end{eqnarray}
We desire an expression for the energy valid to second order in $A_q$.  To that end we Taylor expand Eq.~(\ref{CoulInt}) with respect to $r$ up to second order and then sum up the contributions from each pair of charges $n$ and $m$.  This is a straightforward but lengthy procedure and we quote only the result.  Throwing out negligible terms which do not scale with the number of charges in the system, we obtain the change in potential energy
\begin{eqnarray}\label{SHelixEnergyInt}\nonumber
\lefteqn{\delta V/|A_q|^2 =\sum_{n,m} -4 (1 - \cos q h(n-m))\cos(n-m)\psi }\\ \nonumber
&&\times \int_0^{\infty} (\frac{w}{\pi})^{1/2}e^{-w(h^2(n-m)^2 + 2 r^2(1-\cos(n-m)\psi))}dw \\\nonumber
&&+ 8(1-\cos q h(n-m))(\sin(n-m)\psi)^2 \\ 
&&\times \int_0^{\infty} (\frac{w^3}{\pi})^{1/2}e^{-w(a^2(n-m)^2 + 2 r^2(1-\cos(n-m)\psi))}dw.\nonumber \\
\end{eqnarray}
Note that the resulting expression depends on the indices $n$ and $m$ only through their difference.  This means that a sum over $n$ with $n-m$ fixed yields a multiplicative factor going as the number of charges in the helix.  Thus, the potential energy per charge is given by summing over $k=n-m$ in Eq.~(\ref{SHelixEnergyInt}).  This sum was re-expressed in a form suitable for quick numerical evaluations using the Ewald summation approach, just as in the helix-helix interaction calculation shown in the appendix.  The procedure is again straightforward.  However, the resulting expressions are lengthy and provide little intuition and so will not be presented here.  Two plots of the resulting dispersion relations are given in Fig.~\ref{1mode:fig}.  In each case the initial angle between charges was chosen to be $\psi=\pi$.  In the first case $h=3.0 r$ and in the second case $h=1.8r$.  What we see in the first plot looks like a typical phonon dispersion relation.  As the wavevector increases, the frequency increases from zero.  In the second plot, however, we see that there is no real frequency solution for small wavevectors.  This indicates that the $\psi=\pi$ system is unstable with respect to global rearrangements of the charges for this value of $h$.

The instability of the $\psi = \pi$ system at small $h/r$ values can be understood through a consideration of the energy of a single helix of charge.  Because the self-energy becomes variable for flexible systems, this energy must also be taken into account when considering inter-helical interactions.  In order to calculate this energy for an isolated, perfect helix, we once again carry out an Ewald summation.  We sum up the contributions from each charge to the potential at a given point on the cylinder.  We then take the limit as this point approaches the position of one of the charges and subtract off the interaction with that charge to get the potential due to each of the other charges at this location.  The divergence of the potential due to the infinite number of other charges on the cylinder is canceled out when added to the potential from the neutralizing charge distribution up the center of the helix.  The resulting expression for the energy per charge is

\begin{eqnarray}\nonumber
\lefteqn{E = \frac{\sqrt{\beta}}{h} (-2 + \sum_{n \not =0}\int_1^{\infty} e^{-(2 r^2(1-\cos \psi n)+(h n)^2)\frac{\beta \pi}{h^2}t }t^{-1/2}dt) }\\ \nonumber
&&+\int_0^{\pi \beta/a^2}\frac{(I_0(2 r^2 t)e^{-2 r^2 t} - e^{-r^2 t}) }{h t} dt-\int_{\pi \beta/h^2}^{\infty}\frac{e^{-r^2 t}}{h t}dt \\ \nonumber
&& \\ 
&&+\frac{1}{h}\sum_{l,m'}  \int_0^{\pi \beta/h^2}I_l(2 r^2 t) e^{-2 r^2 t - \frac{(2 \pi m + \psi l)^2}{4 h^2 t}} t^{-1}dt.
\end{eqnarray} 
In the above, $\beta$ is the constant determining the cut-off between the high and low integration domains, the $I_l$ are Bessel functions, and the primed sum is over all $l$ and $m$, excluding the term where they are both zero.

\begin{figure*}
\subfigure[]{
\includegraphics[scale=0.9]{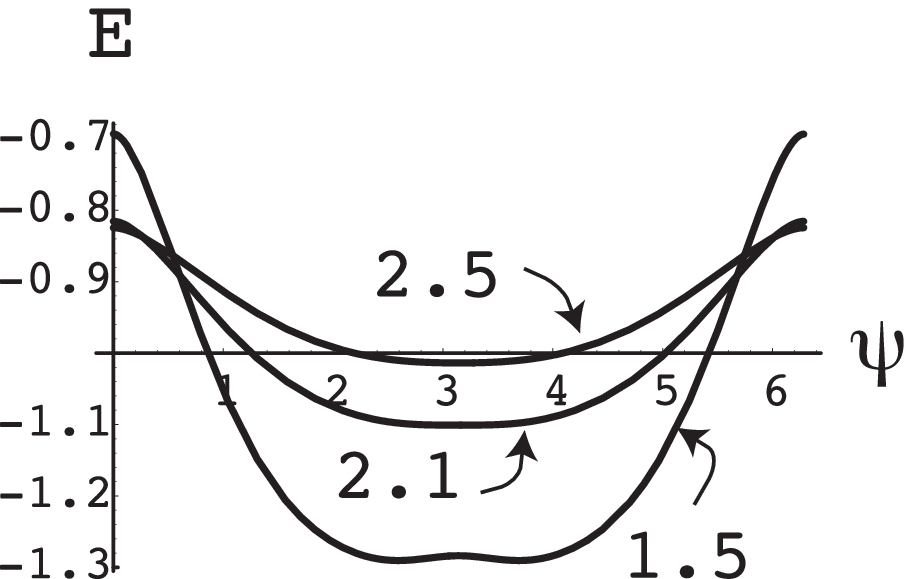}
}
\subfigure[]{
\includegraphics[scale=0.9]{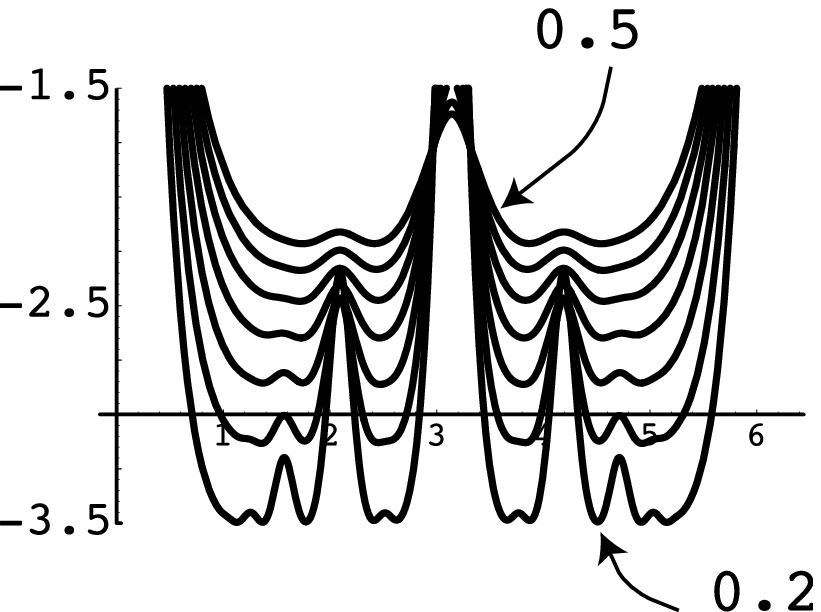}
}
\caption{\label{1energy:fig} Plots of the single helix energy as a function of $\psi$ for the various values of $h/r$ indicated in the figures.}
\end{figure*}

Resulting energy versus $\psi$ plots for a single helix are shown in Fig.~\ref{1energy:fig}.  At large values of $h/r$ a single energy minimum appears at two charges per turn.  However, as the value of $h/r$ decreases below approximately $2.1$, a bifurcation occurs and two new minima replace the original minimum at $\psi = \pi$.  A slight twist is introduced, either right or left-handed, in order to increase the separation distance between axially adjacent charges on each side of the cylinder.  This explains the observed instability of the $\psi = \pi$ system at small $h/r$.  As $h/r$ is further decreased, more and more nearly degenerate minima appear.  This is explained below.

\begin{figure*}
\subfigure[\label{miller:fig1}]{
\includegraphics[scale=0.30]{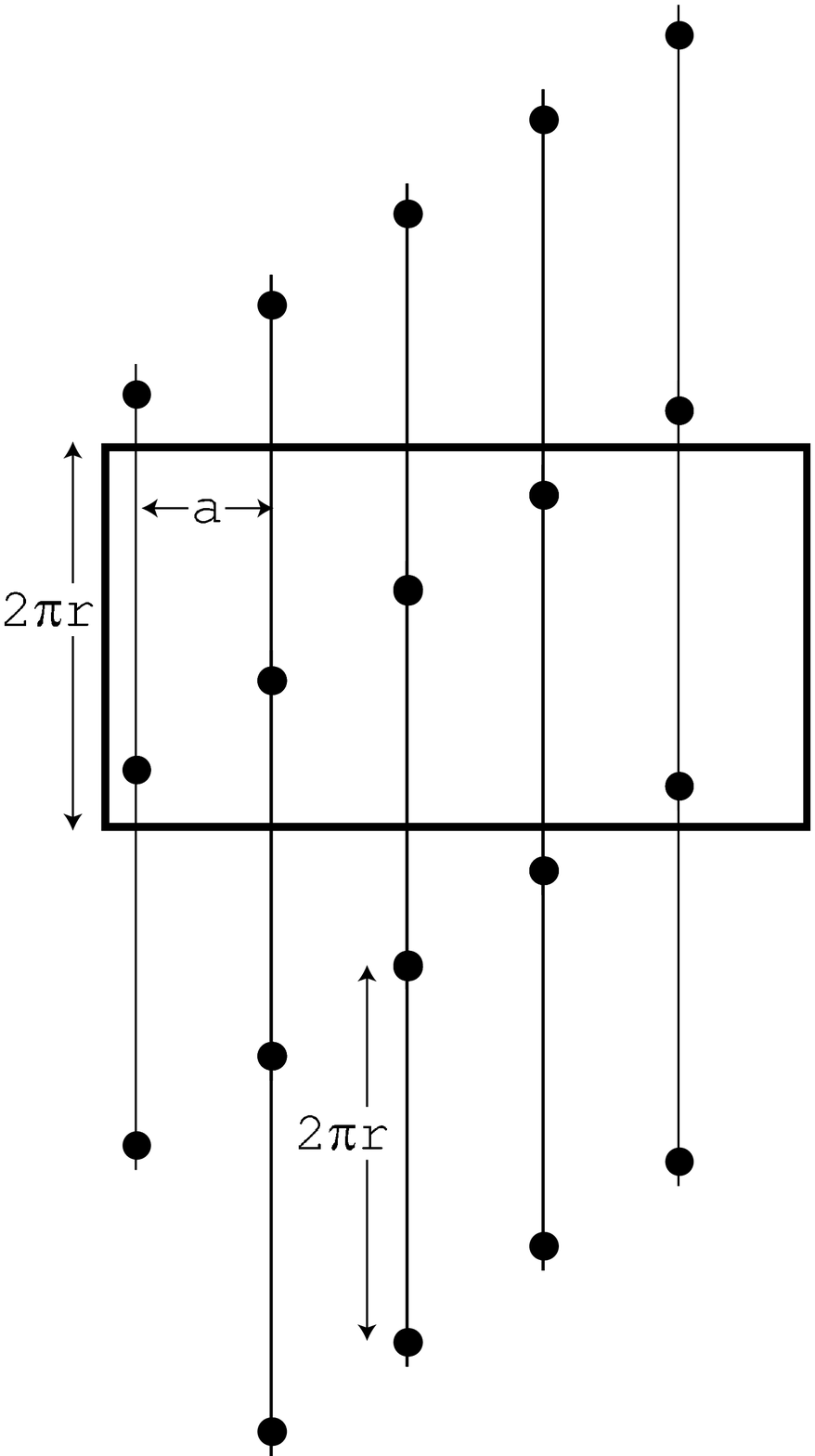}
}
\subfigure[\label{miller:fig2}]{
\includegraphics[scale=0.8]{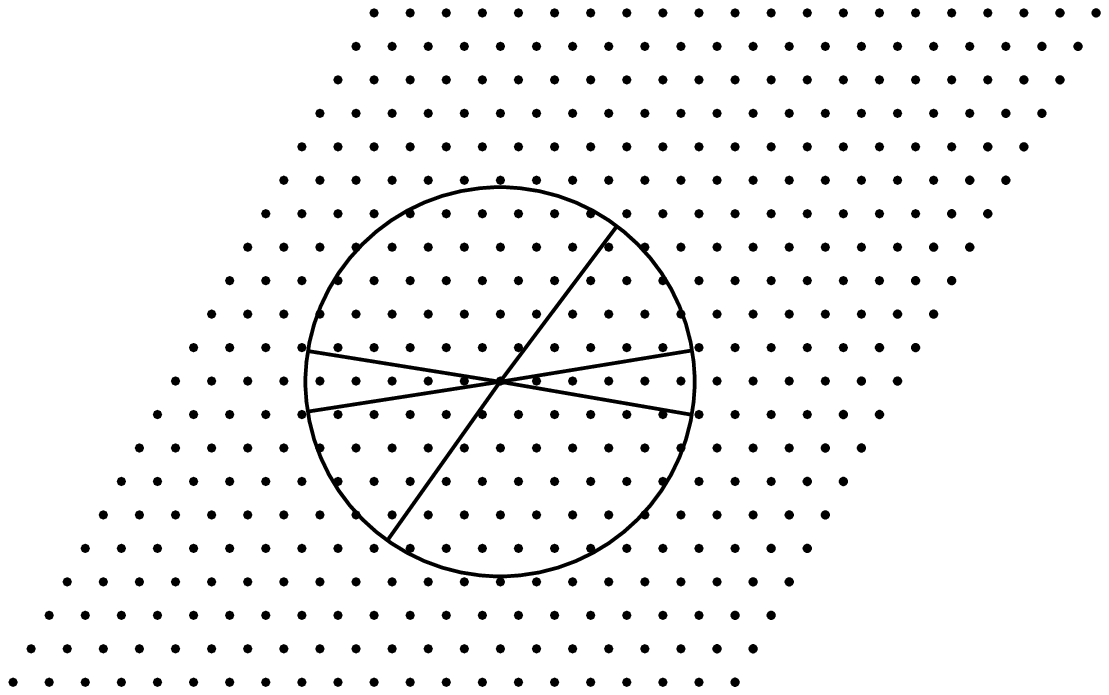}
}
\caption{\ref{miller:fig1} The spiral arrangement of charges on the rolled out cylinder.  \ref{miller:fig2}  Candidates for the Miller line.  }
\end{figure*}

\begin{figure*}
\subfigure[]{
\includegraphics[scale=0.75]{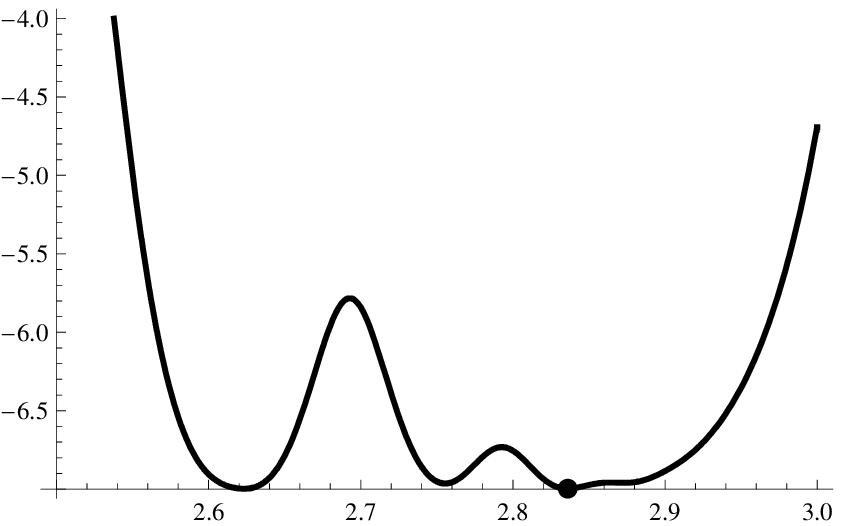}
}
\subfigure[]{
\includegraphics[scale=0.9]{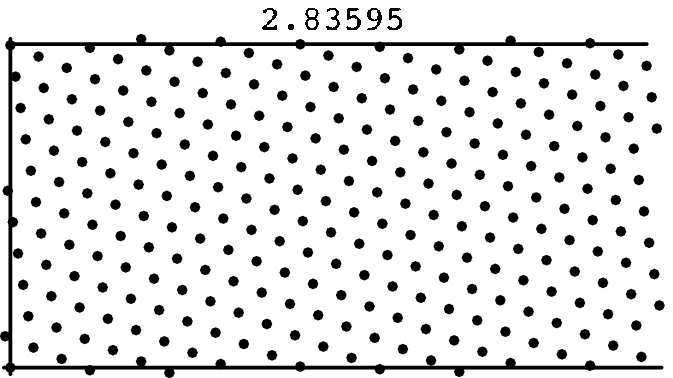}
}
\caption{\label{exlattice:fig} Shown are the energy minimum at $\psi=2.83595$ which appears at $h/r = 0.05$ and the rolled out lattice structure it corresponds to.}
\end{figure*}

At small $h/r$, the charges are tightly packed onto the surface of the cylinder.  They will thus attempt to arrange themselves into a structure resembling an energy minimizing triangular lattice.  To determine which $\psi$ values allow for nearly triangular lattice packings, consider what the charges would look like if we were to unroll the cylinder.  What we would see is what appears in the rectangle in Fig.~\ref{miller:fig1}.  The rectangle is a portion of the rolled out cylinder, which continues to the left and right.  The charges on the cylinder, viewed in this way, are a portion of a Bravais lattice of charges in which the spacing between vertical lines along which the charges lie is $h$ and the distance between charges on one of the lines is $2 \pi r$, as shown in the figure.  We may interpret the lines on which the charges lie as Miller lines of the lattice.  Note that the volume of the primitive cell of this lattice, $v_{WS} = 2 \pi r h$, is completely determined by $r$ and $h$.  Consider now a triangular lattice with this same primitive cell volume.  Suppose this triangular lattice has a Miller line on which the charges are spaced by $2 \pi r$.  This would determine a helical configuration for our cylinder of radius $r$ and axial rise $h$ which would unroll into this triangular lattice structure.  The angle $\psi$ between charges would be determined by the spacing between charges on adjacent Miller lines.  In general, there will probably be no Miller line on the triangular lattice with charge spacing $2 \pi r$.  However, there may be Miller lines with spacings which are close to $2\pi r$.  If this is the case, a slight distortion of the triangular lattice would result in a possible structure for the helix which is energetically favorable.  To determine candidate values, one can take a point on the triangular lattice and draw a circle of radius $2 \pi r$ about this point, as  shown in Fig.~\ref{miller:fig2}.  At small $h/r$, the radius will be large compared to the charge spacing and many points will appear near the circle's outer perimeter.  Such points provide candidate Miller lines which upon a slight distortion will determine a helical configuration of charges that results in a nearly triangular lattice.  These indeed correspond to the energy minimizing $\psi$ values.  An example is shown in Fig.~\ref{exlattice:fig}.  We can thus understand both the small and large $h/r$ limits of the single helix energy landscape.

We now consider the interaction between a pair of flexible helices.  Once again we assume that the axial positions of the charges are fixed but allow them to independently rotate in the azimuthal direction.  All charges interact via a Coulomb force law and all charge values are taken to be $1$.  In addition, an elastic energy is introduced which resists the adjustment of the azimuthal separations between nearest neighbor charges on each helix.  The energy cost of adjusting a nearest neighbor separation is $\kappa \Delta \theta^2/2$.  Here, $\kappa$ is an elastic constant and $\Delta \theta$ is the difference between the assumed angular separation for a nearest neighbor pair and $\psi$, the equilibrium angular separation assumed when the helices are isolated.  

In order to determine the mode structure of a given pair of helices, one must first find the equilibrium orientation.  This equilibrium orientation is characterized by a given set of global parameters, but also requires a determination of the optimal internal twistings within each helix.  We employ the matrix version of the Newton-Raphson method to determine these equilibria.  We start by assuming a given orientation for the two helices.  This is specified by the parameters $\zeta$, $\phi$, $\Delta \psi$, and $\psi$, where $\psi$ is again taken to be $2 \pi m_{\psi}/n_{\psi}$.  Periodicity is enforced every $n_{\psi}$ charges for each helix.  Thus, once the distance between the helices and the axial positions of the charges are set, the geometry of the system is determined by the vector $\vec{\theta}$, which contains $2 n_{\psi}$ components specifying the angular positions of the charges on the two helices.  The force vector and the Hessian matrix for the system is then calculated.  Their components are defined as

\begin{eqnarray}
f_i(\vec{\theta}) &=& -\frac{\partial E}{\partial \theta_i} \\
H_{i,j}(\vec{\theta}) &=&  \frac{\partial^2 E}{\partial \theta_i \partial \theta_j},
\end{eqnarray}
where $E$ is the total electrostatic and elastic energy for the system.  If the force vector does not vanish identically, the angular positions of the charges are adjusted by a small amount $\delta \vec{\theta}$ in order to reduce this force.  Assuming the system is near an equilibrium point, one can take a first order Taylor series for the force and set it equal to zero to obtain

\begin{eqnarray}
H_{i,j} \delta \theta_j - f_i \approx 0.
\end{eqnarray}
An approximation to the ideal $\delta \vec{\theta}$ is then obtained after a matrix inversion.  This is added to $\vec{\theta}$ to obtain a new $\vec{\theta}$ and the process is repeated until it converges to an equilibrium location.

\begin{figure}
\includegraphics[scale=0.8]{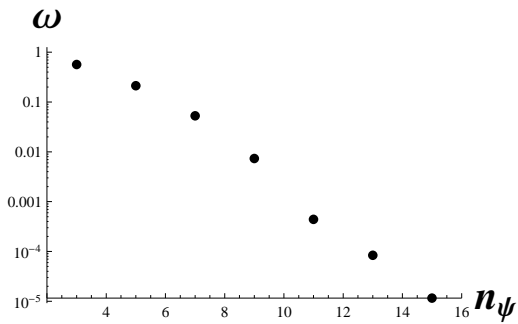}
\caption{\label{fig:pairmodes}  A plot of the oscillation frequency of the symmetric mode versus $n_{\psi}$. In this mode the helices rotate about their respective axes in the same direction.  Here, the system parameters were defined by $d = 2.5 r$, $m_{\psi}=1$, $a=1/\psi = 2 \pi /n_{\psi}$, and $\zeta = 0$.}
\end{figure}

In order to determine the mode structure of the system one need only calculate the eigenvectors and eigenvalues of the Hessian matrix of the energy.  Note that this is conveniently already determined after iteration of the Newton-Raphson method.  Of particular interest are the two lowest energy modes of the system.  These modes involve the rotations of the two helices about their respective axes with little internal degrees of freedom excited.  Indeed, in the rigid limit, these are the only two modes allowed since the relative axial positions of the charges are held fixed.  In one of these modes the helices rotate in the same azimuthal direction while in the other they rotate in opposite directions.  For symmetric equilibrium orientations, the amplitudes of the motions of the two helices are equal for each of these modes.  In this case, the first mode corresponds to adjusting only the $\phi$ global degree of freedom.  From our symmetry arguments above, we know that the interaction energy of an irrational pair of rigid helices does not depend $\phi$.  We thus expect there to be a gap between the irrational and rational frequencies of oscillation in this mode.  In order to test whether or not this could be observed in flexible systems we computed the oscillation frequencies of this mode for various integer number of charges per turn systems.  The elastic constant was taken here to be rather large such that little internal motion would occur.   A log plot of the resulting oscillation frequency versus $n_{\psi}$ is shown in Fig.~\ref{fig:pairmodes}.  Note that the log plot is approximately linear, consistent with the exponential decay in $n_{\psi}$ of the $\phi$ dependence for rigid helices.  We conclude that for systems with low flexibility, a gap should be observed between irrational and rational helices at long wavelength.  

\begin{figure*}
\subfigure[]{
\includegraphics[scale=0.75]{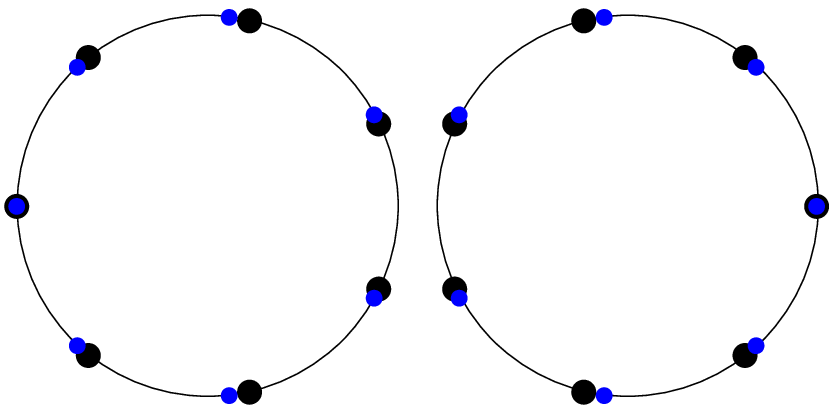}
}
\subfigure[]{
\includegraphics[scale=0.75]{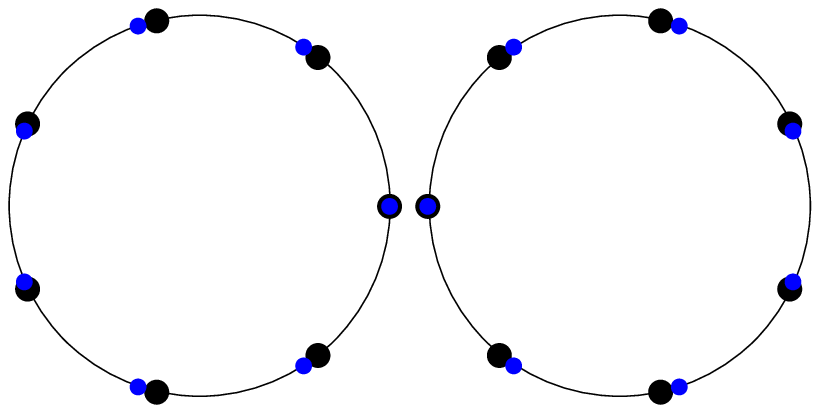}
}
\caption{\label{fig:flex} Shown are end-on views of two equilibrium orientations for a seven charge per turn pair.  The larger black dots represent the locations where the charges would sit for a rigid system at equilibrium and the smaller blue dots represent the corresponding locations for the flexible system.  For a rigid system, the first orientation would be stable.  However, for elastic constants below $\kappa=9$, the value for the system depicted here, the first orientation becomes unstable due to local twisting and the second orientation shown becomes the stable equilibrium.  Here $d=2.2 r$, $\psi = 2 \pi/7$, $a = 1/\psi$, and $\zeta = 0$.  Note that the two charges which appear to be very close to each other in the second figure are actually separated by a significant axial distance.  (Color online) }
\end{figure*}

As the flexibility of these systems is increased local twisting will start to become more significant.  The two helices may then be modelled as a pair of perfect helices plus a series of physical dipoles.  If the dipole strengths are weak, or if the system is in an orientation where the rigid helix azimuthal energy dependence is large, the perfect helix interaction will dominate.  However, as the flexibility is further increased, dipole-monopole interactions can begin to dominate the azimuthal energy dependence.  An example is provided by the seven charge per turn system depicted in Fig.~\ref{fig:flex}.  At and below a threshold value of $\kappa$, the stable equilibrium orientation switches to the second orientation shown.  This corresponds to a change in the equilibrium $\phi$ value of $2 \pi/7$.  This example demonstrates that local twisting can alter the interaction in a qualitatively considerable way even for systems which contain only a modest degree of flexibility.

\section{Physical Applications: F-actin and A-DNA in aggregate}
Both F-actin and A-DNA have been observed to condense under the influence of multivalent counterions.  These two systems occupy different limits in the energy landscape and provide relevant systems to which we may apply our effective energy expressions above.  In deriving these expressions, we assumed that the interacting charges formed perfect helical distributions.  However, these same expressions also often apply for disordered, flexible systems.  To prove this, one must note that the pair interaction is invariant under the operation
\begin{eqnarray}\label{pair:symmetry3}
(\phi, \Delta \psi, \zeta) \rightarrow (-\phi, -\Delta \psi, -\zeta).
\end{eqnarray}
This operation is equivalent to rotating the helices by $180$ degrees so that they are flipped upside down.  This symmetry, together with that in Eq.~(\ref{pair:symmetry}), allows for an immediate, and more general, derivation of the phase relationships observed in Eqs.~(\ref{energy:zeta}), (\ref{energy:deltapsi}), and (\ref{energy:rational}).  This proof applies whenever the variables $\phi$, $\Delta \psi$ and $\zeta$ are well-defined and provide a sufficient characterization of a pair's mutual orientation.  In particular, it applies if there is uncorrelated, non-additive disorder in the charge locations and it also applies if the system is flexible.  Note that flexibility introduces many more degrees of freedom to a system.  However, there should remain a single well-defined ground state orientation for any choice of the three parameters $\phi$, $\Delta \psi$ and $\zeta$, and it is the energy of this ground state which will be represented by the effective energy expressions considered here.  Note that this is consistent with the change in equilibrium orientation observed in the flexible system depicted in Fig.~\ref{fig:flex}; apparently increasing the flexibility in a system can allow for the amplitudes of the various terms to change signs, but the phases must remain fixed.  

The argument outlined above demonstrates that the effective energy expressions derived in section IV apply whenever the pair interaction maintains the periodicities $\phi$, $\Delta \psi$ and $\zeta$.  The effective energy expressions will thus allow us to model many physical aggregate systems without having to know the detailed form of the interaction.   There is one caveat, however; the interaction energy in a system of flexible helices is no longer pair additive, due to the fact that helices may interact indirectly through the local twisting of third party helices.  In the applications below, we assume that the pair interaction dominates such effects.  This assumption may not hold for highly flexible systems.

\subsection{F-actin}
The average $\psi$ value in Mg$^{2+}$ condensed paracrystals of F-actin is $2 \pi 6/13 \approx 2.9$ radians.  This value differs from the average $\psi$ value observed for isolated actin filaments by about $0.5$ degrees per monomer \cite{Ege:82}. 
Numerics indicate that for a discrete $\psi = 2 \pi 6/13$ helix pair, with radii and $a$ values scaled appropriately for actin, the continuous energy dominates globally.  That is, the energy is minimized at, or very near $\zeta = 0$.  Further, at $\zeta = 0$ it is found that the irrational energy terms dominate the rational energy terms and the extremum at $\Delta \psi = 0$ is the location of the energy minimum.  This is consistent with the observation that filaments within a given layer of these paracrystals are typically found to be in register \cite{Fow-82}.  Taking $\zeta = \Delta \psi = 0$ to be the location adopted by the physical system, it follows from Table \ref{table1} that the amplitude of the first $\phi$ component correction term is zero.  This will have the effect of greatly reducing the energy's dependence on the parameter $\phi$.

Numerics also suggest that the irrational $\psi$ dependence cannot alone be responsible for the observed twisting of F-actin in aggregate.  For small $d$, the energy minimum occurs at $\psi = \pi$, which would always tend to increase $\psi$.  This is inconsistent with the observation that different types of actin were each observed to twist to $\psi = 2 \pi 6/13$, regardless of whether the isolated $\psi$ values were less than or greater than this value \cite{Ege:82}.  At larger $d$ values, near $d\approx 5 r$, a moving energy minimum appears near this value.  At these distances the energy gain associated with the irrational energy is insignificant compared to the energy cost due to twisting, which was evaluated using the torsional rigidity constant reported in Ref.~\cite{Tsu:96}.
It thus appears unlikely that charge discreteness effects can allow for sufficient energy gains to explain the $0.5$ degree twist per monomer observed in Ref.~\cite{Ege:82}.  

More recently, tightly packed Ba$^{2+}$ condensed F-actin bundles have been observed which have been torsionally twisted to an average $\psi$ value of $2 \pi 17/36$ \cite{Ang:03}.  Note that the results of Table \ref{table1} indicate that the first $\phi$ component would not vanish for a $\zeta = \Delta \psi =0$ hexagonally packed lattice of molecules of this configuration.  The value $n_{\psi} = 36$ suggests that the rational correction terms should be highly damped for this system, however.  Nevertheless, without knowing the precise form of the interaction we cannot rule out that this twisting was induced in order to obtain a rational energy benefit.  Actin monomers are highly heterogeneous, and as the authors of Ref.~\cite{Ang:03} point out, their highly-charged subdomain-1s may often dominate the interactions.  

It is important to point out that F-actin is particularly flexible in the azimuthal direction.  It is thus possible that the inclusion of non-pairwise additive local twisting effects might help to explain the observed averaged deformation angles.

\subsection{A-DNA}

Previous numerical work has shown that for both A-DNA and B-DNA the correction terms due to discreteness cannot be mutually optimized for each nearest neighbor pair in a hexagonally packed lattice \cite{Kor:98:1}.  Due to frustrations in the $\zeta$ dependence, B-DNA is typically observed to pack in an orthorhombic lattice, however \cite{Lan:60}.  The resulting reduction in the number of nearest neighbors for each molecule makes it more likely that frustrations in the rational dependence can also be overcome.  As discussed below, there is no $\zeta$ frustration in aggregates of A-DNA.  However, A-DNA is also not observed to form hexagonally packed aggregates \cite{ful:65}.  In order to strengthen the plausibility of the suggestion in Ref.~\cite{Kor:98:1} that other lattice structures are assumed in order to reduce azimuthal frustration, we shall now revisit the problem of packing A-DNA on a hexagonal lattice.  We find that to lowest order the frustration results in a complete averaging out of the dominating azimuthal, discrete energy terms.  

Because A-DNA is a double helix, we must consider four single-helix pair interactions for each pair of neighboring molecules.  Each of the single helices of A-DNA is observed to have 11 charges per turn in aggregate.  Experiments and numerics indicate that the energy is optimized when there is no axial shift between the two molecules \cite{Kor:98:1}.  This implies that two of the single-helix pairs will correspond to $\zeta =0$ interactions.  The remaining two interactions, which correspond to the upper helix of one molecule interacting with the lower helix of the other, will have non-zero $\zeta$ values.  At all $\zeta$ values the rational energy terms are observed to dominate the irrational energy terms and we may approximate the azimuthal energy dependence by the first order rational correction to the energy.  Because the energy is periodic in $\Delta \psi$ with period $\psi$, we may expand the amplitude of the first $\phi$ component,  $A_{1,\phi}$, as
\begin{eqnarray}
A_{1,\phi}= \sum_{k=0}^{\infty} \alpha_k \sin \bigg(\frac{11(2k+1)\Delta \psi}{2} + \gamma_k \bigg).
\end{eqnarray}  
Numerics indicate that the first term alone often accurately models the interaction and we have the following approximate expression for the pair interaction's azimuthal dependence.
\begin{eqnarray}
\label{adna-energy}
E_{\phi} = \alpha \sin(\frac{11\Delta \psi}{2} + \gamma) \cos 11(\phi + \frac{\pi - \Delta \psi}{2}).
\end{eqnarray}
Here, both $\alpha$ and $\gamma$ are functions of $\zeta.$

Consider first the interactions between the upper helix strands.  For these interactions $\zeta=0$ and the results of Table \ref{table1} indicate that $\gamma =0 $.  Plugging in Eq.~(\ref{adna-energy}) for each of the six nearest-neighbor interactions for a given molecule demonstrates that the energy is independent of that molecule's azimuthal orientation.  This implies that these terms average out in the bulk.  For the same reason the interactions between the lower helix strands must also average out.

Now consider the two non-zero $\zeta$ interactions for a given pair of molecules.  If one of the interactions corresponds to $\zeta = \zeta_{UD}$ the other must correspond to $\zeta = \zeta_{DU} = -\zeta_{UD}$.  Here the subscript $UD$ indicates the interaction between the upper strand of the first molecule and lower strand of the second, while the subscript $DU$ indicates the opposite interaction.  To relate the $\alpha$ and $\gamma$ values for these two interactions we apply the transformation appearing in Eq.~(\ref{pair:symmetry}) which leaves the energy invariant.
\begin{eqnarray} \nonumber
\lefteqn{\alpha(\zeta_{UD})\sin [\frac{11 \Delta \psi}{2} + \gamma(\zeta_{UD})]\cos 11(\phi + \frac{\pi - \Delta \psi}{2}) }\\ \nonumber
&&= \alpha(\zeta_{DU})\sin [\frac{11 \Delta \psi}{2} - \gamma(\zeta_{UD})]\cos 11(\phi  + \frac{\pi - \Delta \psi}{2}).  \\
\end{eqnarray}
It follows that
\begin{eqnarray}
\alpha(\zeta_{UD}) &= &\alpha(\zeta_{DU}) \\
\gamma(\zeta_{UD}) &=& - \gamma(\zeta_{DU}).
\end{eqnarray}

One final geometric effect must be taken into account.  In general, the azimuthal positions of the charges on the upper strands may be shifted with respect to the positions on the lower strands.  Let the mean shift be $\delta$ so that for two corresponding charges on the upper and lower strands of one molecule we have, $\bar{\theta}_{i,U} = \bar{\theta}_{i,D} + \delta$.  With this definition the sum of the two non-zero $\zeta$ interaction terms is
\begin{eqnarray}\nonumber
\lefteqn{E_{\phi,UD}+E_{\phi,DU}} \\ \nonumber
&=& \alpha \bigg \{ \sin [\frac{11 (\Delta \psi - \delta)}{2} + \gamma] +\sin [\frac{11 (\Delta \psi + \delta)}{2} - \gamma] \bigg \} \\
&& \cos 11(\phi + \frac{\pi - \Delta \psi - \delta}{2}) \\
&=& 2 \alpha \cos(\frac{11 \delta}{2}-\gamma) \sin \frac{11\Delta \psi}{2} \cos 11(\tilde{\phi} + \frac{\pi - \Delta \psi}{2}),
\end{eqnarray}
where $\tilde{\phi} = \phi - \delta/2$.  This has the same form as the $\zeta=0$ interactions and, therefore, also averages out in the bulk.

In the Kornyshev-Leikin theory of helix-helix attraction, a large number of counterion charges are assumed to bind to the grooves of the helices \cite{Kor:97:1}.  This groove binding allows for the correlations necessary for attraction.  To take any groove bound charges into account, we assume, as in Ref.~\cite{Kor:97:1}, that the groove bound charges are not azimuthally ordered, and so taken together may be modelled as continuous helices of charge.  The resulting lowest order expression for the interaction energy between one phosphate strand of one A-DNA molecule and a continuous, condensed counterion helix on a second A-DNA molecule takes the form $A \cos 11(\phi + \delta)$.  This energy form averages out when summed over the six nearest neighbors of each molecule in a hexagonally packed lattice.  Thus, taking all of the interactions into account, we have seen that to lowest order, the azimuthal energy terms completely average out in a hexagonally packed A-DNA system.  This provides a strong statement regarding the degree of frustration for this geometry and is consistent with the idea that non-hexagonal packing structures are adopted in order to overcome azimuthal frustration.

\section{Discussion}
The work presented here is complimentary to that presented in Refs.~\cite{Kor:98:1} and \cite{Kor:98:2}.  The model considered in these previous articles attempted to incorporate the effects of counterions explicitly.  Here, we have focused directly on the symmetries of the interaction and have obtained some results which are, in a sense, model independent.  In particular, the effective energy expressions derived can be applied to model both electrostatic and non-electrostatic aspects of the pair interaction.

In our characterization of the helix-helix interaction we began by decomposing the energy into a sum of terms, each of which added dependence to the energy on a new parameter.  For Coulomb interactions, the rational energy correction term was shown to decay exponentially with $n_{\psi}$.  This result can also be shown to hold for any power law or Yukawa interaction.  This characterizes how rational a helix pair has to be in order to obtain a significant rational energy benefit.  Symmetry arguments allowed us to demonstrate the existence of locations where the dominating rational correction term vanishes.  The exact location in parameter space where this occurs was determined for certain high symmetry orientations of the two helices.  Finally, the phases of the Fourier series expansions were determined for each of the correction terms.  

We next considered the mode and energy structure of an isolated flexible helix of charge.  When the axial shift per charge is greater than $h\approx 2.1$, the single helix energy is minimized when there are two charges per turn.  Slightly below $h \approx 2.1$, axially adjacent charge interactions cause the energy to be minimized when the helix is slightly twisted either to the right or to the left.  A series of similar energy bifurcations were observed to occur as $h$ was further decreased.  It follows that the energy of a single helix can be highly dependent upon twist angle.  This may often play a role in determining the equilibrium conformation of helical molecules.  For interacting pairs of helices, the $\phi$ independence of the irrational energy suggests that a gap should occur between rational and irrational systems at long wavelength.  Scattering experiments could thus, in principle, provide information regarding the degree of rationality in a system of bundled helices.  Although the phases of the interaction terms remain the same for flexible systems, it was shown numerically that the amplitudes can change sign.  This means that the ground state orientation for a pair of helices can change drastically as flexibility is increased.   

Although we did not consider the general problem of discrete frustrations in aggregate systems, the two applications we covered demonstrate that such studies may be carried out easily on a case by case basis.  For $13/6$ charges per turn F-actin paracrystals, our findings appear to indicate that the observed twistings in aggregate cannot be explained by the electrostatics of the pair interaction.  Given the recently presented results indicating that twisting may limit bundle width in protein linked F-actin aggregates \cite{Cla:08}, it seems plausible that local twisting may play some key role in the mechanics of counterion-induced aggregation of F-actin as well.  For A-DNA, our brief consideration of a hexagonally packed system indicated that the dominating rational term completely averages out for this geometry.  This further strengthens the plausibility that non-hexagonal packing structures are observed for A-DNA to reduce this frustration.

We conclude with a few comments on local twisting.  First, we note that an irrational-rational transition is still expected for flexible systems, though the boundaries of the phase diagram may be changed dramatically depending on the stiffness of the system.   As shown above, an infinitesimal twisting to a rational system can allow for a finite energy gain while costing a negligible amount of torsional twisting energy.  Thus rational transitions are still expected to often occur.  In highly flexible systems, however, local discommensuration states are allowed and compete with the rational, fully in register, states.  At finite temperature, the entropic gain associated with such states may make them more favorable.  Second, we reiterate that the pair interaction studied in this paper may not accurately model highly flexible systems.  For aggregates of DNA, we expect the pair interaction to be accurate, as DNA is known to have an unusually high torsional rigidity \cite{Bau:97}.  Actin on the other hand is known to have an especially low torsional rigidity \cite{Ege:82}, implying that corrections to the pair interaction might be significant in this case, as mentioned above.

\begin{acknowledgements}
The authors thank Professor Robijn Bruinsma for helpful comments.
\end{acknowledgements}

\appendix*

\section{The Coulomb sum}
In this appendix we outline the steps taken to express the perfect helix interaction energy in a form which may be easily evaluated numerically.  Formally, the energy is given by Eq.~(\ref{E:sum:def}), which we rewrite below.
\begin{eqnarray}
\label{E:def:app}
E =  \sum_{\theta_1, \theta_2} \frac{\exp[-a_s R]}{R}.
\end{eqnarray}
The sum here is over all pairs of charges, one taken from each helix, where, from Eq.~(\ref{R:def}), $R$ is given by  
\begin{eqnarray}\nonumber
R^2 &=&
d^2 - 2rd[\cos(\theta_1 - \phi) - \cos(\theta_2 - \phi)] \\ 
&&+ 2r^2[1- \cos(\theta_1 - \theta_2)] + a^2[\theta_1 - \theta_2 + \zeta]^2.\nonumber \\
\end{eqnarray}
To begin we apply the identity
\begin{eqnarray}
 \frac{\exp[-a_s R]}{R} = \pi ^{-1/2}\int_0^{\infty} t^{-1/2}\exp[i k a_s - (k^2 + R^2)t] dt, \nonumber \\
\end{eqnarray}
and then replace the exponentiated cosines in (\ref{E:def:app}) using the series representation
\begin{eqnarray}
\exp[y \cos \theta] = \sum_{l=- \infty}^{\infty} I_l (y) \exp[i l \theta].
\end{eqnarray}
Here $I_l$ is a modified Bessel function of the first kind.  Then, the expression that will yield the interaction energy is
\begin{eqnarray} \nonumber
\pi^{-1/2} \int_0^{\infty} t^{-1/2} I_{l_1}(2drt) I_{l_2}(-2drt) I_{l_3}(2r^2t) \\ \exp \bigg[ -\frac{a_s^2}{4t}  -t \bigl[d^2 + 2r^2   + a^2( \theta_1 - \theta_2 + \zeta)^2 \bigr] \nonumber \\+il_1(\theta_1 - \phi) + il_2( \theta_2- \phi) + il_3 ( \theta_1 - \theta_2) \bigg] dt.  \nonumber \\
\label{eq:int4}
\end{eqnarray}
We now let $n_1$ and $n_2$ in Eqs.~(\ref{n1:def}) and (\ref{n2:def}) be given by
\begin{eqnarray}
n_1 &=& m_a + m_b  \\
n_2 &=& m_b.
\end{eqnarray}
The sum on $m_b$ may then be separated out.  It is
\begin{eqnarray}\nonumber \lefteqn{
\sum_{m_b = -\infty}^{\infty} \exp[i(l_1 + l_2)m_b \psi] }\\&=& 2 \pi \sum_{k=-\infty}^{\infty}\delta(\psi(l_1+l_2) - 2 \pi k). 
\label{eq:msum2}
\end{eqnarray}
If $\psi/2 \pi$ is irrational, then the only possible way in which an argument of one of the delta functions on the right hand side of Eq.~(\ref{eq:msum2}) can be zero is if $l_1 + l_2 =0$. For the time being, we will assume that this is the case. Then, we have an infinite contribution from all terms of the form $l_1=-l_2$. To see what happens in this case, we set $l_1=-l_2$ on the left hand side of Eq.~(\ref{eq:msum2}). The result is that we are summing over one for each charge on one of the helices. The reason that we end up with an infinite result is the sum as defined in Eq.~(\ref{eq:msum2}) contains an infinite number of terms. If we are interested in the energy per charge, we simply take one of them, having set $l_1+l_2=0$. This leaves us with the following sets of sums and integrals to perform
\begin{eqnarray}
\lefteqn{\sum_{l_1,l_3,m_a=-\infty}^{\infty} \int_{0}^{\infty} dt \  ( \pi t)^{-1/2}I_{l_1}(2rdt) I_{-l_1}(-2rdt) I_{l_3}(2r^2t)} \nonumber \\ && \exp \bigg[-\frac{a_s^2}{4t}-t(d^2+2r^2)-a^2t(m_a \psi - \Delta \psi + \zeta)^2  \nonumber \\&& +i(l_1+l_3)( m_a \psi - \Delta \psi) \bigg ].
\label{eq:int5}
\end{eqnarray}

We now split the range of integration over $t$ into two parts.  When $t>1/(a \psi)^2$ there is reasonably rapid damping of the sum over $m_a$.  Let's call the lower limit for this range of $t$ values $T$.  For the upper range of $t$ values, we invert the sum over $l_3$ to get
\begin{equation}
\sum_{l_3=-\infty}^{\infty} I_{l_3}(2r^2t) e^{il_3(m_a- \Delta \psi)} = \exp[2r^2t \cos(m_a- \Delta \psi)].
\label{eq:l3sum}
\end{equation}
Through application of the identity
\begin{equation}
I_n(x) = \frac{1}{2 \pi} \int_0^{2 \pi} e^{-in \theta + x \cos \theta},
\label{eq:bf1}
\end{equation}
the sum on $l_1$ may be evaluated to give
\begin{eqnarray}
\lefteqn{\sum_{l_1=-\infty}^{\infty} I_{l_1}(2rdt)I_{-l_1}(-2rdt) e^{il_1(m_a - \Delta \psi)}} \nonumber \\&=&
\frac{1}{2 \pi}  \int_0^{2 \pi} \exp [2rdt( \cos \theta  \nonumber \\ \nonumber
&&-\cos( \theta - m_a \psi + \Delta \psi + 2 \pi k)] d \theta \\
\label{eq:bf3}
\end{eqnarray}
Reconstructing the expression, we are left with the following integral and double sum
\begin{eqnarray}
\lefteqn{\frac{1}{2 \pi} \sum_{m_a=-\infty}^{\infty} \int_0^{2 \pi} d \theta \int_T^{\infty} dt \   ( \pi t)^{-1/2} \nonumber  \exp \bigg[  -\frac{a_s^2}{4t} }\\&& -t(d^2+2r^2)    -a^2 \psi^2 t(m_a-(\Delta \psi - \zeta)/\psi)^2  +2rdt [ \cos \theta   \nonumber \\ && - \cos( \theta - m_a \psi + \Delta \psi )] +2r^2 t \cos(m_a \psi - \Delta \psi) \bigg]. \nonumber \\
\label{eq:larget1}
\end{eqnarray}
It is important to note that because of the damping of the sum over $m_a$ the number of terms actually summed over is not large. Furthermore, the integration over $\theta$ is limited to a finite range. Finally, the integration over $t$ is exponentially convergent. 

In the small-$t$ regime, we concentrate on the sum over $m_a$ in Eq.~(\ref{eq:int5}). Here, we use the following identity, based on the Poisson sum formula
\begin{equation}
\sum_{m=-\infty}^{\infty}f(m) = \int_{-\infty}^{\infty} f(m) \sum_{k^{\prime}=-\infty}^{\infty} e^{2 \pi i k^{\prime} m} dm.
\label{eq:poiss1}
\end{equation}
This leads to the following expression to be summed and integrated over in the low-$t$ regime. 
\begin{eqnarray}
\lefteqn{\sum_{l_1,l_3,k^{\prime}=-\infty}^{\infty} \int_0^T dt \frac{1}{a \psi t} I_{l_1}(2drt) I_{-l_1}(-2drt) I_{l_3}(2r^2t)} \nonumber \\ &&  \exp \bigg[ - \frac{a_s^2}{4t}  - t(d^2+2r^2)  -\frac{( 2 \pi k^{\prime} +(l_1+l_3) \psi)^2}{ 4 a^2 \psi^2 t} \nonumber  \\
&& -i \zeta (l_1+l_3) + i\frac{ \Delta \psi - \zeta}{\psi} 2 \pi k^{\prime}  \bigg]. \nonumber \\
\label{eq:smallt2}
\end{eqnarray}
This last transformation has ensured that the sum on $k^{\prime}$ is now quickly damped in Eq.~(\ref{eq:smallt2}).

Things get a bit simpler when the interaction is unscreened, in that at least one of the integrals above can be expressed in terms of special functions. However, in the absence of screening the energy per charge diverges as expressed in Eq.~(\ref{E:def:app}).  This divergence may be eliminated if we subtract out the energy per charge for two interacting lines of charge.  We may then add this term back on using the logarithmic expression for the energy, obtained by shifting the origin of the first line's potential.  The term we subtract out is
\begin{equation} 
\frac{1}{a \psi} \int_0^{\infty} t^{-1} \exp \bigg[-d^2 \ t \bigg]  dt,
\label{eq:counter5}
\end{equation}
which diverges in the small-$t$ limit.  The appropriate arrangement of canceling terms is the following:
\begin{eqnarray}\nonumber
\lefteqn{\mathcal{C}(d,r,T) }  \\ &\equiv & \nonumber
\frac{1}{a \psi} \int_0^T \frac{1}{t} \bigg\{ I_0(2drt)^2 I_0(2r^2t) \exp \left[ -t(d^2+r^2)\right]  \\ 
&& -\exp[-d^2t]\bigg\} \ dt.
\label{eq:counter6}
\end{eqnarray}

The simplification that results from removing the screening is in the integration over the upper limit. What we can do is express the integral over $t$ in terms of an exponential integral or of an error function. The relevant results are
\begin{eqnarray}
\int_T^{\infty}\frac{e^{-wt}}{t}dt &=& - \mathop{\rm Ei}(-Tw)
\label{eq:expint1} \\
\int_T^{\infty}t^{-1/2}e^{-wt}dt & = & \frac{\sqrt{\pi } \mathop{{\rm erfc}}(\sqrt{T w})}{\sqrt{w}}.
\end{eqnarray}

The final expression for the interaction between two irrational helical arrays of charge, for the integration from 0 to $T$ is
\begin{eqnarray}
\lefteqn{\sum_{l_1,l_3,k^{\prime}=-\infty}^{\infty} (1- \delta_{l_1^2+l_3^2+k^{\prime \, 2}}) \int_0^T dt \frac{1}{a \psi t} I_{l_1}(2drt)} \nonumber \\ && I_{-l_1}(-2drt) I_{l_3}(2r^2t)  \exp \bigg[  - t(d^2+2r^2)  \nonumber \\ && -\frac{( 2 \pi k^{\prime} +(l_1+l_3) \psi)^2}{ 4 a^2 \psi^2 t} -i \zeta (l_1+l_3) + i\frac{ \Delta \psi - \zeta}{\psi} 2 \pi k^{\prime}  \bigg] \nonumber \\
&& +  \mathop{\mathcal{C}}(d,r,T) - \frac{2}{a \psi} \log d.
\label{eq:fullsmallt}
\end{eqnarray}
And for the integration from $T$ to $\infty$, we have
\begin{eqnarray}
\lefteqn{\frac{1}{2 \pi} \sum_{m_a=-\infty}^{\infty} \int_0^{2 \pi} \bigg[d^2+2r^2 +a^2 \psi^2 (m_a-(\Delta \psi - \zeta)/ \psi)^2} \nonumber \\ && -2rd[ \cos \theta - \cos( \theta-m_a \psi + \Delta \psi)] \nonumber\\&& -2r^2 \cos(m_a \psi - \Delta \psi) \bigg]^{-\frac{1}{2}}   \mathop{\rm erfc}\bigl(T^{1/2} [d^2+2r^2 \nonumber \\ &&+a^2 \psi^2 (m_a-(\frac{\Delta \psi - \zeta)}{\psi})^2  -2rd[ \cos \theta \nonumber \\ &&
- \cos( \theta-m_a \psi + \Delta \psi)]  -2r^2 \cos(m_a \psi - \Delta \psi) ]^{1/2} \bigr) d \theta \nonumber \\ && + \frac{1}{a \psi} \mathop{\rm Ei}(-d^2T)
\label{eq:fulllarget}
\end{eqnarray}
Although the expressions in Eq.~(\ref{eq:fulllarget}) are a bit cumbersome, their numerical evaluation, at least in \emph{Mathematica}, is considerably easier.

For rational $\psi = 2 \pi \frac{m_{\psi}}{n_{\psi}}$, the delta function generated by the summation over $m_b$ in Eq.~(\ref{eq:msum2}) is a bit more general. Now the condition is
\begin{equation}
l_1+l_2 = 2 \pi k_{\psi} n_{\psi}
\label{eq:newdel}
\end{equation}
where $k_{\psi}$ is also an integer. After a bit of work, we find for the new version of Eq.~(\ref{eq:fulllarget})
\begin{eqnarray}
\lefteqn{\frac{1}{n_\psi} \sum_{m_a=-\infty}^{\infty} \sum_{l_{\psi}=1}^{n_{\psi}}\bigg[d^2+2r^2 +a^2 \psi^2 (m_a-(\Delta \psi - \zeta)/ \psi)^2} \nonumber \\ && -2rd[ \cos \Theta(m_a, l_{\psi}, \phi) - \cos( \Theta(m_a, l_{\psi}, \phi)-m_a \psi + \Delta \psi)] \nonumber \\ && -2r^2 \cos(m_a \psi - \Delta \psi) \bigg]^{-1/2} \nonumber \\ &&  \mathop{\rm erfc}\bigl(T^{1/2} [d^2+2r^2 +a^2 \psi^2 (m_a-(\Delta \psi - \zeta)/ \psi)^2 \nonumber \\ && -2rd[ \cos \Theta(m_a, l_{\psi}, \phi) - \cos( \Theta(m_a, l_{\psi}, \phi)-m_a \psi + \Delta \psi)] \nonumber \\ && -2r^2 \cos(m_a \psi - \Delta \psi) ]^{1/2} \bigr)  +  \frac{1}{a \psi} \mathop{\rm Ei}(-d^2T),
\label{eq:fulllargetrat}
\end{eqnarray}
where 
\begin{equation}
\Theta(m_a, m_{\psi},\phi)= m_a \psi - \phi + \frac{2 \pi l_{\psi}}{n_{\psi}}.
\label{eq:thetafun}
\end{equation}
Finally, the new version of Eq.~(\ref{eq:smallt2}) is
\begin{eqnarray}
\lefteqn{\sum_{l_1=-\infty}^{\infty} \sum_{l_3=-\infty}^{\infty} \sum_{k=-\infty}^{\infty} \sum_{k^{\prime}=-\infty}^{\infty}  \int_0^T
dt \frac{1}{a \psi t}I_{l_1}(2drt)I_{-l_1 +kn_{\psi}}(-2rdt)   } \nonumber \\
&& I_{l_3}(2r^2t) \exp \bigg[ -\frac{a_s^2}{4t} -t(d^2+2r^2) - \frac{(2 \pi k^{\prime}+(l_1+l_3) \psi )^2}{4a^2 \psi^2 t}  \nonumber \\ && - i \zeta(l_1+l_3) + i \frac{\Delta \psi - \zeta}{\psi} 2 \pi k^{\prime} + i k n_{\psi}( \Delta \psi - \phi) \bigg].
\end{eqnarray}

\bibliography{refs}

\end{document}